\RequirePackage{ifpdf}    
\documentclass[hyper]{JHEP3} 

\pdfoutput=1
\usepackage{graphicx}
\usepackage{epsfig}
\usepackage{amsmath}
\usepackage{cite}
\usepackage{lscape}
\usepackage{multirow}
\usepackage{color}
\let\normalcolor\relax
\makeatletter

\newcommand{\fmslash}[2][0mu]{%
  \mathchoice
    {\fmsl@sh\displaystyle{#1}{#2}}%
    {\fmsl@sh\textstyle{#1}{#2}}%
    {\fmsl@sh\scriptstyle{#1}{#2}}%
    {\fmsl@sh\scriptscriptstyle{#1}{#2}}}
\newcommand{\fmsl@sh}[3]{%
  \m@th\ooalign{$\hfil#1\mkern#2/\hfil$\crcr$#1#3$}}
\makeatother

\newcommand{\beq}{\begin{equation}}
\newcommand{\eeq}{\end{equation}}
\newcommand{\bea}{\begin{eqnarray}}
\newcommand{\eea}{\end{eqnarray}}

\newcommand{\lsim}{{\;\raise0.3ex\hbox{$<$\kern-0.75em\raise-1.1ex\hbox{$\sim$}}\;}}
\newcommand{\gsim}{{\;\raise0.3ex\hbox{$>$\kern-0.75em\raise-1.1ex\hbox{$\sim$}}\;}}
\newcommand{\met}{\not \!\! E_T}
\newcommand{\mptvec}{\not \!\! \vec{P}_T}

\title{Constrained $\sqrt{\hat{S}_{min}}$ and reconstructing with semi-invisible production at hadron colliders}

\author{Abhaya Kumar Swain and Partha Konar \\
	     Theoretical Physics Group, Physical Research Laboratory (PRL),\\
             Ahmedabad, Gujarat - 380 009, India \\
            E-mail: \email{abhaya@prl.res.in}, \email{konar@prl.res.in}
            }

\received{\today}               
\accepted{\today}               

\preprint{\today}

\abstract{ 
Mass variable $\sqrt{\hat{S}_{min}}$ and its  variants~\cite{Konar:2008ei, Konar:2010ma} were constructed
by minimising the parton level center of mass energy that is consistent with all inclusive measurements. 
They were proposed to have the ability to measure mass scale of new physics in a fully model independent way. 
In this work we relax the criteria by assuming the availability of partial informations of new physics events 
and thus constraining this mass variable even further. Starting with two different classes of production topology, 
{\it i.e.} antler and non-antler, we demonstrate the usefulness of these variables to constrain the unknown masses. 
This discussion is illustrated with different examples, from the standard model Higgs production and 
beyond standard model resonance productions leading to semi-invisible production. We also utilise these constrains 
to reconstruct the semi-invisible events with the momenta of invisible particles and thus improving the 
measurements to reveal the properties of new physics.
}


\begin{document}

\section{Introduction}
\label{sec:1}

The Standard Model (SM) is now essentially complete after CMS~\cite{Chatrchyan:2012ufa} and 
ATLAS~\cite{:2012gk} found its last missing bit, lone neutral scalar of the model, the Higgs boson. 
The SM is so far extremely successful in explaining the fundamental particles and the interactions 
between them. However some of the unresolved theoretical questions together with very convincing 
experimental observations, such as dark matter, neutrino oscillation and several others compel 
us to believe that the SM can not be the complete description. Numerous models beyond Standard 
Model (BSM) was constructed to accommodate some of these phenomena with a general belief that 
the scale of new physics is just around the corner at few to multi-TeV level. Unfortunately, 
large hadron collider (LHC) has not observed any indication of new physics so far.
Now, if any of these TeV scale BSM theories exists in nature then it can manifest its signature 
at the next LHC run. A scenario with positive signal essentially necessitates the determination 
of the new particle mass, spin and coupling etc associated with that new physics. 

Recently popular theoretically appealing BSM theories are the ones which accommodate the 
thermal relic dark matter as stable and weakly interacting massive particle (WIMP) 
estimating the tightly constrained observed amount of dark matter density~\cite{Hinshaw:2012aka}. 
Hence, this stability of the dark matter in most of the BSM theory is ensured by some discrete 
symmetry, such as $Z_2$ symmetry in supersymmetry or many other scenario. Once this symmetry is 
respected, all the heavy BSM particles in such model has to be produced in pairs; subsequently 
decaying into some lighter BSM resonance together with SM particles (which may or may not be 
detected and measured at the detector) in multiple steps of successive  decay. Typically at the end of 
each decay chain lightest BSM particle is produced which is the dark matter particle 
of that model and escape the detection. Hence, at least two massive and lightest BSM particles 
remain hidden in these events. The only way 
to know their presence is the observation of sizable $\mptvec$ in the detector calculated from the 
imbalance of transverse visible momenta produced in such events. The reconstruction of a dark matter 
signal at hadronic collider is challenging because of the partial knowledge of the incoming parton 
momenta further burdened with multiple massive final state particles of unknown mass
goes undetected keeping no individual momentum informations at the detector.

There has been several studies under gone into  mass and spin determination in the context 
of semi-invisible production at the  hadronic collider\footnote{For some recent review, 
see ref.~\cite{Barr:2010zj, Barr:2011xt, Kong:2013xma}} 
and we classify them based on the topology information as follows:

\textit{Exclusive variables} are defined 
based on the topology of the production mechanism and decay processes under consideration. 
Identical signatures consists of visibles and invisibles in the final state can be originated from very different topologies
which is deeply related to the stabilising symmetry of the dark matter (DM). Shape of the visible 
invariant mass can effectively carry informations on topology along with the mass spectrum~\cite{Cho:2012er} of the decay chain.
Underlying DM stabilising symmetry can also be probed~\cite{DMstabilizingKinematicEdge, DMstabilizingMT2, DMstabilizingCountinginvisible}
using kinematic edge and cusp in the invariant mass distributions and from the shapes of transverse mass variable $M_{T2}$. 
Even the assumption of one particular underlying symmetry allows some fixed number of different topologies
from which the correct one can be identified comparing suitable kinematic variables~\cite{identifyDMTopology}. 
One expects that the ignorance of the 
correct topology can add difficulties in solving combinatorial ambiguity~\cite{reducingcombinatoricsMT24bjet, Rajaraman:2010hy, Baringer:2011nh, reducingCombinatoricsMT2andMAOS} which is
one source of complexity in mass determination methods, more prominently available when associated with long decay chain. 
This ambiguity can be originated from two different
sources. Firstly, allocation of the final state particles to the correct decay chain, {\it i.e.} from
which side of the decay chain some particular states is produced. Secondly, the ordering of 
the assigned particle in a single decay chain. 
The hemisphere method~\cite{Ball:2007zza} and PT vs M methods~\cite{Rajaraman:2010hy} are introduced to 
reduce the this ambiguity in assigning the correct final state particles to the corresponding decay chain. 
However, the ordering of the particles left unresolved. The $M_{T2}$ variable together with invariant mass  
are also shown to reduce the combinatorics significantly~\cite{Baringer:2011nh}.
In the literature several classes of exclusive variables are defined assuming that the correct knowledge 
of topology is available and anticipating that the combinatorial ambiguity can be controlled. 
The exclusive mass determination methods can be categorised  as follows
\begin{itemize}
 \item \textit{Edge measurement method}: Based on the idea of constructing all possible invariant masses 
 out of visible decay products in each decay chain~\cite{edgeMesurement, edgeMeasurementambiguity,  edgeMeasurementambiguityLow, edgeMeasurementmassAmbiguity, edgeMeasurementexptResolution, edgeMassmeasurementmassRatio, edgeMeasurementmassRatio2}. 
 Each invariant mass has an endpoint which is experimentally observable and these endpoints are related 
 to the unknown masses in the decay chain. To evaluate all the unknown masses by inverting the 
 the equations in terms of measured endpoints, one needs sufficient number of independent endpoint measurements. 
 So essentially a long decay chain in necessary to have unique measurement of all the 
 unknown masses. However this criteria inevitably invites combinatorial ambiguity thereby reducing the 
 effectiveness of the method. This method also does not use all the available informations like missing 
 transverse momentum $\mptvec$ in the event.
 \item \textit{Polynomial method}: One tries to utilise all the available information in the event of 
 particular topology and solve for the unknown masses and momentums~\cite{polynomialmethod, polynomialmethod1, polynomialmethod2, polynomialmethodtop, polynomialmethodaccuratereco} 
 considering on-shell cascade decay. In the literature, typically the production of two heavy invisible particle is considered 
 at the final state, assuming $Z_2$ type of DM stabilising symmetry in the theory.
 All the unknown invisible momenta components are  
 solved utilising mass-shell constraints and missing $\mptvec$ constraints in the event. It can be shown that
 one needs to consider long decay chains to solve for all unknowns in the event. 
 Combinatorial ambiguity naturally arises here from the requirement of the long decay chain. Moreover, 
 resulting invisible momenta remain ambiguous due to existence of multiple 
 solutions originating from non-linear mass-shell constraints~\cite{AnalyticalSol2,polynomialmethodaccuratereco}.

 \item \textit{Transverse mass variable}: Rather than considering full event information, 
 transverse projection of momenta is considered during calculation. Contrary to previous cases, even 
 small decay chain can constrain the masses realistically.  
 There are many variants of transverse mass variable exist in the literature such as $M_{T2}$~\cite{MT2first, MT2TruthbehindGlamour, 
 TopPartnersSpinAndMassMeasurement, MTGEN, MT2Kinkprl, MT2KinkJHEP, MT2KinkWithExtraPT, MT2Kink110WihPT, MT2Hemisphere}, $M_{T2}^{sub}$~\cite{MT2Subsystem},
 $M_{CT2}$~\cite{Cho:2009ve, Cho:2010vz}, 1D orthogonal decomposition of $M_{T2}$ ($M_{T2\perp}$ and $M_{T2\parallel}$)~\cite{Konar:2009wn},
 asymmetric $M_{T2}$~\cite{Barr:2009jv, Konar:2009qr} and $M_{T2}^{approx}$~\cite{Lally:2012uj}, $M_{CT}$~\cite{Tovey:2008ui, Polesello:2009rn, Serna:2008zk},
 and variants $M_{CT\perp}$ and $M_{CT\parallel}$~\cite{Matchev:2009ad} etc.
 Among these broad class of transverse mass-bound variables, 
 we briefly discuss some properties of $M_{T2}$ which is studied widely in the literature.
 This variable is defined  as the constrained minimisation of maximum of two transverse masses $M_T$ from both
 sides of the decay chain. The minimisation is done over all 
 possible partitions of missing transverse momenta where as, satisfying the $\mptvec$ constraint. 
 $M_{T2}(\tilde{m}_{inv})$ expressed as a function of the unknown invisible particle mass, can have an 
 experimentally observed upper bound over many events. This provides a useful correlation between the trial 
 invisible mass $\tilde{m}_{inv}$ and measured upper bound $M_{T2}^{max}$, which represents the corresponding mass of the 
 ancestor particle (commonly called as mother or parent) responsible for producing all the visible and 
 the invisible particles within the (sub)system. This correlation also satisfy
 the true yet unknown mass parameters fulfilling the crucial equality $M_{T2}^{max}(m^{true}_{inv})=m^{true}_{mother}$.
 Interestingly, one can measure the true mass of both mother and daughter simultaneously by identifying 
 a discontinuity (kink) arises due to additional conditions like two step decay chain~\cite{MT2Kinkprl, MT2KinkJHEP},
 extra upfront $P_T$ from ISR~\cite{MT2Kink110WihPT, MT2KinkWithExtraPT} or in subsystem context~\cite{MT2Subsystem}.
 Extracting these kinematic endpoint is occasionally troublesome with thinly populated events at the endpoint, and in presence of backgrounds. Available on-shell constraints of intermediate particles can be exploited in
 the (1+3) dimensional variable $M_2$~\cite{Mahbubani:2012kx, Cho:2014naa} to improve number the events appearing at the tail of these distributions.

\end{itemize}

\textit{Global and inclusive variable} are independent of topology information and hence, do not require 
any information about the production mechanism of the particles in the event. Variables are constructed 
using only visible particles momenta and missing transverse momenta in the event. Several of them were well known and
utilised for long as event selection variables {\it e.g.} $H_T$~\cite{Tovey:2000wk}, 
total visible invariant mass $M$~\cite{Hubisz:2008gg},  effective mass $M_{eff}$, total transverse component of invisible
momentum $\met$, total visible energy $E$ and total transverse energy $E_T$ in the event.
Lately introduced $\hat{s}_{min}$~\cite{Konar:2008ei} and its variants $\hat{s}_{min}^{sub}$ and 
$\hat{s}_{min}^{reco}$~\cite{Konar:2010ma} are also constructed as global and inclusive variable 
for measuring mass scale of new physics. 
Being a topology independent variable, they are also applicable to any possible decay chain without 
worrying  about the symmetric and asymmetric topology and simple  analytical form is also available.
Thus any generic topology can be assimilated without affected by the combinatorial ambiguity.
We would discuss further about these variables in Sec.~\ref{sec:2}.

In this present work our objective is to demonstrate the usefulness of partial event 
informations including the topology in the variables like $\hat{s}_{min}$ which were 
constructed as global one. After an introduction of analytic form for these variables 
in Sec.~\ref{sec:2}, we discuss effects of additional constrains for two topology classes
based on the production and decay of heavy resonance. 
Antler topology being one important topology for single resonant production,  
have considerable discovery potential in different SM and BSM modes. This class can be 
constrained significantly and some interesting features can be noticed. In Sec.~\ref{sec:3} 
we motivate and discuss the constraints. We would notice the constrained variable 
$\hat{s}_{min}^{cons}$ can notably improve the distribution. Second interesting 
takeaway from these constraints is that not only the minimum quantity $\hat{s}_{min}$, 
but one can also construct a maximum quantity $\hat{s}_{max}$ which is also bounded. Hence, one additional 
variable $\hat{s}_{max}^{cons}$ can be defined and finite since these constraints play 
a critical role. 
After construction of these constrained variables we display how they were restricted in several events.
In the next section, Sec.~\ref{sec:4} we consider similar variables in case of non-antler topology.
In Sec.~\ref{sec:5} we turn to the capability of these variable in reconstructing the events. 
$M_{T2}$ assisted method (known as MAOS)~\cite{MAOS1,MAOSclassification} is proposed earlier by utilising 
the transverse components of the invisible momenta obtaining from the minimisation of this transverse mass.
Whereas, longitudinal components are solved using the on-shell constraints and thus having two fold ambiguity
from each decay chain.
Recently proposed some of the (1+3) dimensional $M_2$~\cite{Cho:2014naa} variables can lead to unique momentum reconstruction
for symmetric topology where additional constraints over equality of mother and on-shell relative take a pivotal role.
Here in the reconstruction from the minimisation of $\hat{s}$ does not rely on any particular topology and can be used for 
any of the symmetric or asymmetric cases for a unique solution. We would further demonstrate that the inclusion of 
additional constraints also improve these reconstructed momenta.
We summarize and conclude at the end.

\section{ $\sqrt{\hat{s}}$ mass-bound variables without additional constraints}\label{sec:2}

Let us start by discussing briefly about the variable $\sqrt{\hat{s}}_{\min}$ which was first introduced~\cite{Konar:2008ei} 
to determine the mass scale associated with any generic process (or event topology) involving missing particles. It is inspired 
from the fact that the precise knowledge of partonic system center of mass (CM) energy $\sqrt{\hat{s}}$ carries the kinematic
informations like masses of heavy resonance, or threshold of pair productions at the hadron collider. Hence, one may like to know the distribution of this variable even approximately, 
after recognising the fact that there is no way we can completely reconstruct the event, 
or extract all the momentum informations in case of general semi-invisible productions at the hadron collider.
Utilising all the experimentally observed quantities, best one gets the minimum
partonic CM energy which is compatible (or consistent) with the observed visible momenta and missing transverse momentum.
Although general event topology can have a wide diversity in production mechanism of visibles and invisibles and also number 
of them, it emerged that the final minimisation leads to a rather simple and versatile functional form for $\sqrt{\hat{s}}_{\min}$.

This variable was further extended~\cite{Konar:2010ma} to apply in general subsystems, and also utilised reconstructed events 
to safe guard the generic variables from underlying events and ISR~\cite{Papaefstathiou:2009hp, Papaefstathiou:2010ru}. 
Subsequently these  $\sqrt{\hat{s}}$ variables were shown and
classified~\cite{Barr:2011xt, Barr:2012im} as $M_1$ type of mass-bound variables 
represented in a compact nomenclature of $M_{\dots}$ class of variables. Wide variety within this class are constructed 
systematically considering different projection methods, additional second projection~\cite{Konar:2009wn, Matchev:2009ad}, 
and considering different orders of the operations. 
Interestingly, most of the existing mass variables devised based on different utility can be 
accommodated in this unified picture, leaving many more new variable elements in this class hitherto unexplored.

One can simplify the discussion under the following assumptions which are rather common in wide class of BSM models:
(i)The DM stabilisation is respected by discrete $Z_2$ symmetry. As a result, all BSM particles in the 
theory would produce in pair leading to two stable DM particles in the final state. They stay invisible
in the detector resulting missing transverse energy as their combined footprint. 
(ii)There is only one DM candidate in the theory, or if multiple DM particles are there then they are degenerate in mass. 
One can note that even after making these two assumption the variable  $\hat{s}_{\min}$ remains global and inclusive.

\begin{figure}[t]
\centering
 \includegraphics[scale=0.6,keepaspectratio=true]{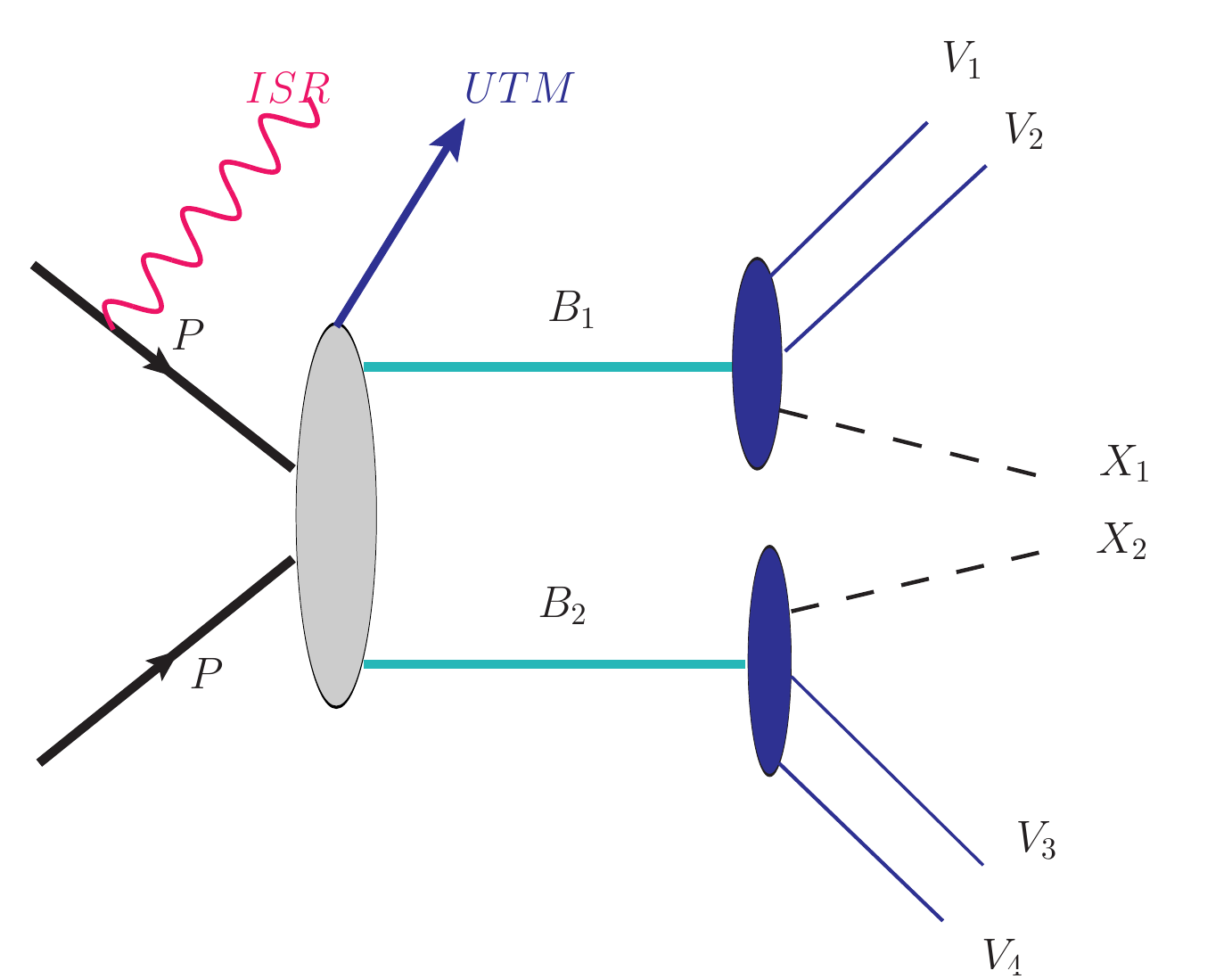}
 \caption{Representative for a simple non-antler topology where after production of two heavy mother particles $B_\alpha$, each of them leading to single invisible massive particle $X_i$ together with number of visibles $V_j$ in the final state. 
 The blue bulb represents the intermediate particle which may be off-shell or on-shell. The visible particles are SM particles measurable at the detector and represented by blue lines denoted by $V_1$, $V_2$, $V_3$ and $V_4$ respectively. The invisible particles are represented by black dashed lines denoted by $X_1$ and $X_2$ respectively.  The momenta of visibles and invisibles are denoted by $p_{i}$, i = 1,2,3,4 and $q_{j}$, j = 1,2.}
 \label{fig:NonAntlerTopology}
 \end{figure}

Under these assumptions analytic expression and properties of this mass-bound variable $\sqrt{\hat{s}}_{\min}$
can be discussed using the non-antler topology displayed\footnote{In general, 
there can be any number of visibles including asymmetric production topology or asymmetric invisibles ({\it e.g.} 
as in~\cite{Konar:2009qr}) in the final state, but here we restrict our discussion for simplicity and as a reference 
for proceeding discussion in following sections. We refer ref.~\cite{Konar:2008ei,Konar:2010ma} 
for a most generic representation for which these $\sqrt{\hat{s}}$ variables are constructed and applicable.} in Fig.~\ref{fig:NonAntlerTopology}. In the (sub)system under consideration,
two mothers denoted by the $B_1$ and $B_2$ either produced in hard scattering at the hadron collider, 
or starting point in the subsystem from a event with longer decay chain. 
Eventually each of these mothers decays to produce two visible and one invisible particle. 
The topology can also contain intermediate particles which may be on-shell or off-shell, 
symbolising into the blue bulb to show the final products only.  Momenta ${p}_{j}$ of these 
visible SM particles $V_j$ ($j=1, \dots ,4$) represented by blue lines can be measured at the detector. On the contrary, 
the invisible particles $X_i$  ($i=1,2$) in black dashed lines are of BSM nature with individual mass $m_i$, and 3-momenta $q_{i}$. The partonic Mandelstam variable for this topology is given by,
\begin{eqnarray}
 \hat{s} = \bigg(E^v + \sum_{i = 1}^{n_{inv}}\sqrt{m_{i}^2 + \vec{q}_{iT}^{\,2} + q_{iz}^2}\bigg)^2 - \bigg(P_z^v + \sum_{i = 1}^{n_{inv}}q_{iz}\bigg)^{\,2}
\end{eqnarray}
Here, $n_{inv} = 2$ is the number of invisible particles, $E^v = \sum_j e^{v}_j$ and 
$P_z^v = \sum_j p^z_{j}$ are  total energy and total longitudinal 
component of the visible momenta. In the above equation missing transverse momentum constraints 
$\mptvec = \sum_i \vec{q}_{iT} $ are also taken into account.
Clearly even in this simplified case, there are $3n_{inv} = 6$ unknown momenta components, 
as well as unknown invisible mass with only two constraints from missing transverse momentum. 
So one can not hope to calculate true values of $\hat{s}$ involved event by event. 
But it is important to realise that there is an absolute minimum exist for $\hat{s}$ in each 
event which also satisfy all these observable. By minimizing $\hat{s}$ with respect to unknown 
momenta $\vec{q}_{i}$ subject to the missing $\mptvec$ constraints one gets
\begin{eqnarray}
 q_{iT} =& f_m^{(i)}  \mptvec \, , \label{eq:qmincon1} \\
 q_{iz} =& f_m^{(i)}  \frac{P_z^v}{\sqrt{(E^v)^2 - (P_z^v)^2}}\sqrt{M_{inv}^2 + \mptvec^2} \, \label{eq:qmincon2}.
\end{eqnarray}
 $f_m^{(i)}$ is a dimensionless mass fraction varies between 0 and 1 and is given 
 by $f_m^{(i)} = \frac{m_i}{M_{inv}}$ and total sum of all invisible masses $M_{inv} = \sum_{i = 1}^{n_{inv}} m_i$. 
 Now replacing the above expression of $q_{iT}$ and $q_{iz}$ in $\hat{s}$ one gets 
 the final form of $\hat{s}_{\min}$ as
 \begin{eqnarray}
  \sqrt{\hat{s}_{\min}(M_{inv})} = \sqrt{(E^v)^2 - (P_z^v)^2} + \sqrt{\mptvec^2 + M_{inv}^2}. 
 \end{eqnarray}
 
 One can follow from the computation that the $\hat{s}_{\min}$ does not assume any particular event topology
 or the DM stabilising symmetry of the model.
 Based on common BSM scenario, we restrict our description (also in Fig.~\ref{fig:NonAntlerTopology})
 assuming $Z_2$ symmetry, so that, pair production of BSM particles are considered producing two invisible massive 
 particles in the final states.
 From minimisation conditions in Eq.~\ref{eq:qmincon1} and~\ref{eq:qmincon2} one can infer that
 each DM particle carries a fraction of missing momenta, proportional to the corresponding mass fraction $f_m^{(i)}$.
 However final $\hat{s}_{\min}$ is simply a function of total $M_{inv}$ irrespective to this fraction.
 Once we assume a pair of invisibles in the final state with same mass (or both massless),
 then this fraction $f_m^{(i)}$ comes out as $1/2$ for any choice of trial mass\footnote{Although, the mass fraction $f_m^{(i)}$ 
 appear to be singular for a choice of zero invisible masses, but one can recalculate starting with a massless scenario and minimizing to 
 get the fraction $ f_m^{(i)} = \frac{1}{2}$. Alternatively, from this present expression with arbitrary masses, one can first use the
 equality of unknown invisible masses before setting it to zero to get back the same fraction.} including the true invisible mass. 
 The invisible momenta at the minimisation are,
 \begin{eqnarray}
  q_{iT} &=&  \frac{1}{2}\mptvec,\\
  q_{iz} &=& \frac{1}{2} \frac{P_z^v}{\sqrt{(E^v)^2 - (P_z^v)^2}}\sqrt{M_{inv}^2 + \mptvec^2}.
 \end{eqnarray}
 
 These invisible momenta calculated from minimisation may not represent that of the true event. However  
 the uniqueness of these momenta can be useful to study the semi-invisible decays involved both in SM as well as BSM scenario.
 Momentum reconstruction can be exploited to analyse the properties of top quark decaying invisibly in the SM, whereas DM 
 motivated BSM models are commonplace where uniqueness of invisible momenta can help to study decays with different topology.
 One can notice that the invisible momenta constructed through $\hat{s}_{\min}$ are always parallel to each other with a magnitude 
 proportional to mass fraction. Here we investigate how the partial knowledge of event information can improve the variable  $\hat{s}_{min}$ and also reconstructed momentum obtained from it. 
 In our further discussion, we divide the production topology as two kinds, such as,  antler topology and non-antler topology. 
 We discuss $\hat{s}_{min}$ with and without putting on-shell constraints in both kinds of these topology.

 \section{Antler topology and constrained variable}\label{sec:3}
 
  Antler topology is very common and well motivated in SM Higgs production. Resonant Higgs production and its semi-invisible 
  decays into W-boson, $h \rightarrow WW^* \rightarrow l \nu l \nu$ or through $\tau$ decay 
  $h\rightarrow \tau \tau \rightarrow w \nu_{\tau}  w \nu_{\tau}$ are some of the interesting channels. 
  Several popular BSM scenario 
  also have these production, such as, supersymmetric (SUSY) heavy Higgs decays through neutralinos
  $H\rightarrow \tilde{\chi}_2^0 \tilde{\chi}_2^0 \rightarrow Z \tilde{\chi}_1^0 Z \tilde{\chi}_1^0$~\cite{Djouadi:2005gj}, 
  or SUSY $Z^{'}$ production with leptonic decay $Z^{'} \rightarrow \tilde{\ell}^{+} \tilde{\ell}^{-} \rightarrow  \ell^{-} \tilde{\chi}_1^0 \ell^{+} \tilde{\chi}_1^0$~\cite{Cvetic:1997wu,Baumgart:2006pa}. In the model of universal extra 
  dimension (UED) one can produce resonant second excitation states decaying into couple of lighter states, like $Z^{(2)} \rightarrow L^{(1)} L^{(1)} \rightarrow \ell^{-} \gamma^{(1)} \ell^{+} \gamma^{(1)}$~\cite{Datta:2005zs,Cheng:2002ej}. 
  Other class of examples being resonant exotics production with their semi-invisible SM decay.
  Doubly charged scalar in hadron collider can decay with one of the dominant decay channel into $w$-pair,
  $\phi^{++} \rightarrow w^{+} w^{+} \rightarrow \ell^{+} \nu_{\ell} \ell^{+} \nu_{\ell}$~\cite{Bambhaniya:2013yca}. 
  Similarly heavy Higgs or heavy $Z^{'}$ can have SM semi-invisible decay 
  $H/Z^{'} \rightarrow t\bar{t} \rightarrow b\bar{b} w^{+} w^{-} \rightarrow b\bar{b} \ell^{+} \nu_{\ell} \ell^{-} \bar{\nu}_{\ell}$. 
  Some of these antler topology was studied~\cite{Han:2009ss,Han:2012nm,Han:2012nr} showing that the invariant mass, 
  transverse momenta and angular correlations constructed out of visible decay products are effective to measure the 
  invisible and intermediate particle masses whereas heavy resonant mass is assumed to be known. 
  Cusp and kink structures appeared in the distributions of these variables
  and their positions are also related to the unknown masses. 
  Missing transverse momentum constraints are not used in these study. 
  Production mechanism in a linear collider ($e^+e^-$) with a fixed center of mass energy is also very similar to 
  this antler decay topology and semi-invisible decays can be studied in similar fashion~\cite{Christensen:2014yya}.

 \begin{figure}[t]
 \centering
 \includegraphics[scale=0.5,keepaspectratio=true]{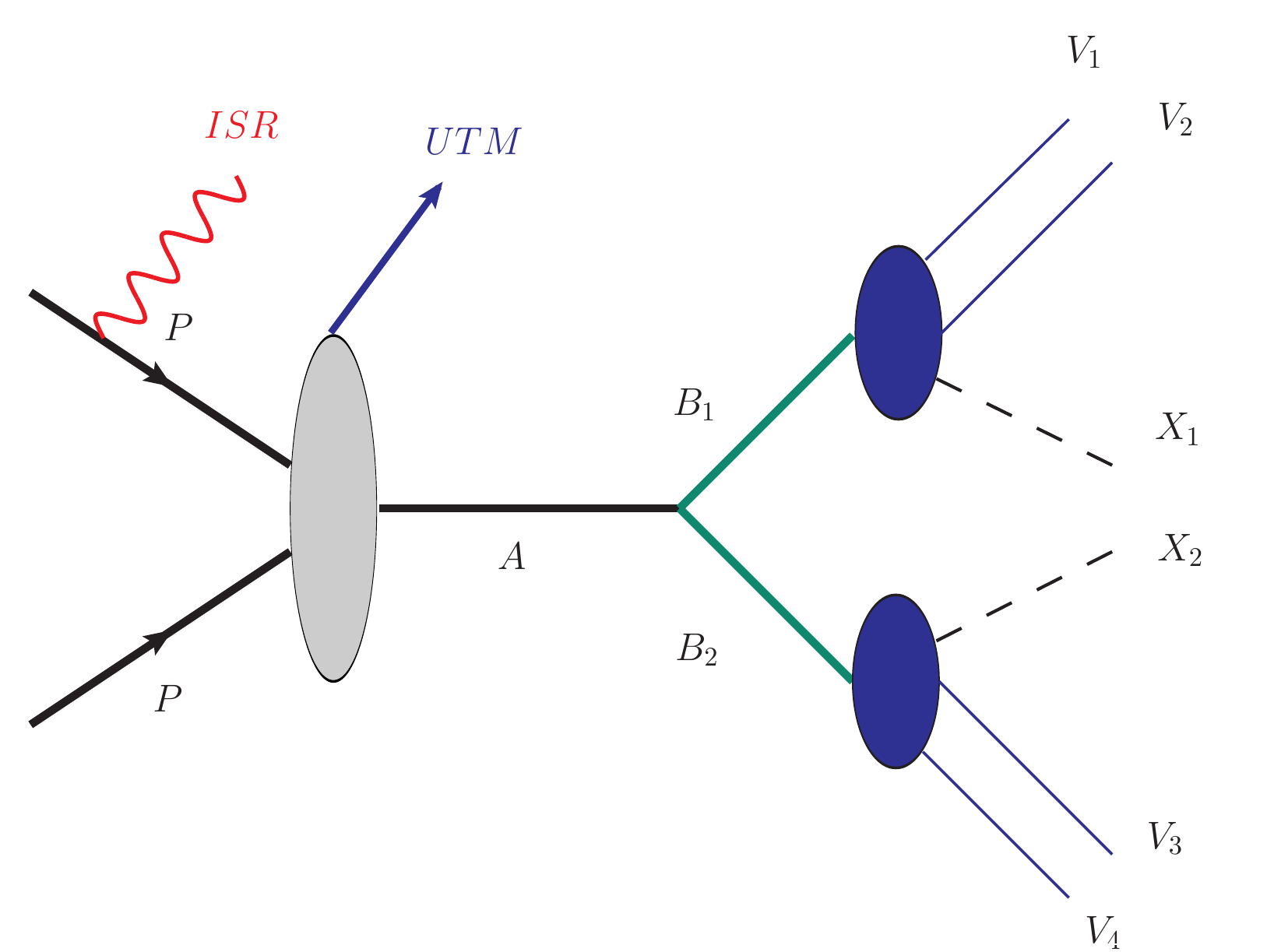}
 \caption{Representative for a simple antler topology where A is a $Z_2$ parity even heavy resonance produces and decays to two daughter particles $B_1$ and $B_2$ and each of which finally decays to two SM visibles $V_j$ and one invisible particle $X_i$.}
 \label{fig:AntlerTopology}
 \end{figure}

 Representative diagram for antler topology is shown in Fig.~\ref{fig:AntlerTopology}.
 Parity even heavy resonant state $A$, produced through on-shell production at the hadron collider, 
 promptly decays to pair of parity odd particles $B_1$ and $B_2$. In this simplified picture, each 
 $B$ subsequently decays same way as we have described earlier in Fig.~\ref{fig:NonAntlerTopology}, and thus
 producing couple of visible with an indivisible daughter. We also keep the same notation for 
 momentum assignment associated to all final particles. 
 Before defining the $\hat{s}_{min}$ in presence of the additional constraints, we first list all the 
 constraints available for this present topology. Apart from the antler resonance mass-shell constraint 
 at some fixed value of the $\hat{s}$ depending upon resonant mass $M_A$,
 \begin{eqnarray}
   (\sum_j{p_j}+\sum_i{q_i})^2 = M_A^2 = \hat{s}^{True},
 \end{eqnarray}
 additional mass equations and missing transverse momentum relations for this topology can be
 put together as, $\{constraints\}$:
 \begin{eqnarray}
  && (p_{1} + p_{2} + q_1)^2 = M_{B_1}^2, \, \, (p_{3} + p_{4} + q_2)^2 = M_{B_2}^2 \, , \label{mother1cons}\\
  && q_1^2 = M_{X_1}^2, \, \, q_2^2 = M_{X_2}^2 \, , \label{daughter2cons}\\
  && \vec{q}_{1T} + \vec{q}_{2T} = \mptvec \label{missingptcons}.
 \end{eqnarray}
 \{$M_{B_1}$, $M_{B_2}$\} and \{$M_{X_1}$, $M_{X_2}$\}  are the masses\footnote{Note that through out the analysis we have 
 assumed both the intermediate and daughter masses are known and used their true masses in the $constraints$.
 However, in a scenario when the invisible particle mass is unknown, one can go ahead with the constrained variables assuming some trial
 mass ($\tilde M_{X1}, \tilde M_{X2}$)  in Eqs.~\ref{daughter2cons}. One can then expect a correlation between this  trial invisible mass with
 the endpoints in constrained variable distributions.} of the intermediate  
 particles \{$B_1$, $B_2$\} and the invisible particles \{$X_1$, $X_2$\} respectively. 
Clearly, using the above constraints in 
 Eqs.~\ref{mother1cons}-\ref{missingptcons} one can reduce the number of free parameter at two. Afterword in Sec.~\ref{sec:5},
we would further demonstrate the constrained regions in these parameter space. One can also notice that in the  
 Eqs.~\ref{mother1cons}  the ordering of the particle in a particular decay chain does not affect the constraints but
 assigning particle to decay chain does. The combinatorial ambiguity of later type is severe when one has long decay chain which is
 absent in our analysis. In general, this type of problem can be partially controlled using exsting methods like hemisphere method~\cite{Ball:2007zza} 
 and PT vs M methods~\cite{Rajaraman:2010hy}.
 
 Now we are in a position to formulate a new variable dubbed as $\hat{s}_{min}^{cons}$ defined as the minimum 
 partonic Mandelstam variable which satisfies all above $constraints$ in the event.
 
 \begin{eqnarray}
  \hat{s}_{min}^{cons} = \min_{\substack{\vec{q_1}, \vec{q_2} \\ \{constraints\}}} [\hat{s}(\vec{q_1}, \vec{q_2})]
 \end{eqnarray}
 Among all constraints defined in the Eqs.~\ref{mother1cons}-\ref{missingptcons}, 
 the variable $\hat{s}_{min}$ is already satisfies last four constraints comprising two 
 missing $\mptvec$ components and two mass-shell constraints from invisible daughters. 
 In other words, new variables are further constrained with mass-shell relations of intermediate parents.

  The true value of partonic Mandelstam variable for antler topology is the mass of the heavy 
  resonance, that is $\sqrt{\hat{s}^{True}} = M_A$, once the heavy resonance produced on-shell and having 
  narrow decay width. Hence, any mass bound variable constructed by minimisation, such as, $\hat{s}_{min}$ 
  for antler topology needs to be bounded from above satisfying the relation $\hat{s}_{min} \le \hat{s}^{True}$.
  This end point can be measured from the endpoint at the distribution over many events.
 Constrained variable $\hat{s}_{min}^{cons}$ also satisfy similar relation  $\hat{s}_{min}^{cons} \le \hat{s}^{True}$,  
 having endpoint at the $\hat{s}^{True}$. However additional intermediate particle mass-shell constraints ensure a 
 larger value of  $\hat{s}_{min}^{cons}$ over $\hat{s}_{min}$ for each event. This inequality would also reflect 
 in the mass variable distributions contributing larger number of events at the endpoint of the distribution.

 As we discussed earlier, a rather striking consequence of this additional mass-shell constraints are that they 
 also permit us to construct a finite upper mass bound variable, which is meaningless otherwise. 
 We define this constrained variable $\hat{s}_{max}^{cons}$ as the maximum partonic Mandelstam variable,
  \begin{eqnarray}
  \hat{s}_{max}^{cons} = \max_{\substack{\vec{q_1}, \vec{q_2} \\ \{constraints\}}} [\hat{s}(\vec{q_1}, \vec{q_2})]
 \end{eqnarray}
 satisfying all the available $constraints$ in the event listed in Eqs.~\ref{mother1cons}-\ref{missingptcons}, which, in turn, is the maximum of the 
 physically allowed region.
 Since $\hat{s}^{True}$ satisfies all the available constraints in the event, it must remain within this region.
 Now, by definition, $\hat s_{max}^{cons}$ is where $\hat s$ is maximum inside this region and 
$\hat s_{min}^{cons}$ is where it is minimum. So, $\hat s^{True}$ can maximally reach up to $\hat s_{max}^{cons}$.
 Hence, $\hat s_{max}^{cons}$ has a lower bound at the $\hat{s}^{True}$,
 significantly with a large number of events at this threshold. 
 Interesting point about these  constrained $\hat{s}$ variables are that the reconstructed momenta from 
 their minimisation(maximisation) not only unique but they also improve over the same calculated through $\hat{s}_{min}$.
 In case of $\hat{s}_{min}^{cons}$, better momentum reconstruction ensured by the points closer to its endpoint. Similarly, 
 $\hat{s}_{max}^{cons}$  gives better reconstruction from the points associated to its threshold. 
 These points would be discussed further in the Sec.~\ref{sec:5}, where these correlations would be more evident. 
 Finally, the definitions of different $\hat{s}$ variables, after imposing different constrains, ensures the hierarchy
 among these mass variables: 
 \begin{equation}
  \hat{s}_{min} \le \hat{s}_{min}^{cons} \le \hat{s}^{True} \le \hat{s}_{max}^{cons}.
 \end{equation}
  
  \begin{figure}[t]
 \centering
 \includegraphics[bb=0 0 360 203,scale=0.55,keepaspectratio=true]{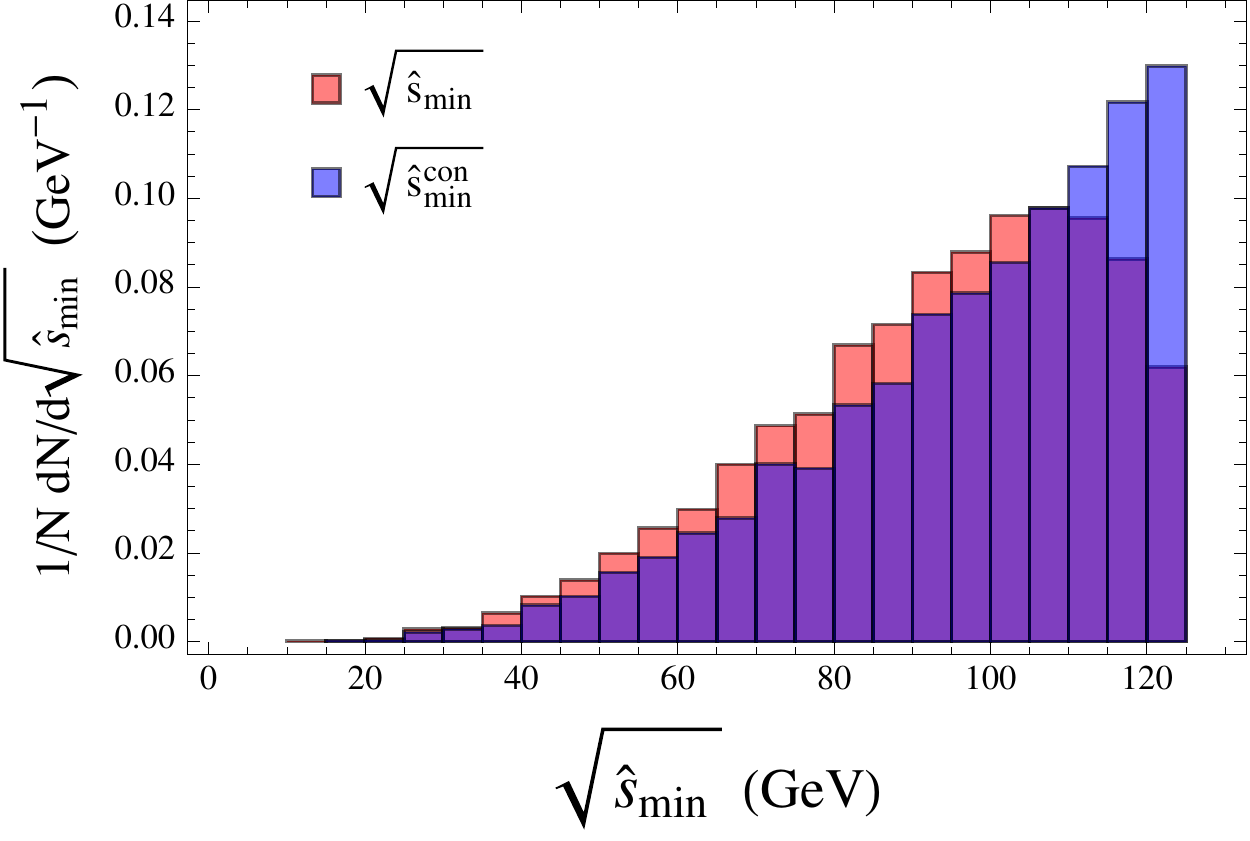}
 \hskip 1cm
 \includegraphics[bb=0 0 360 204,scale=0.48,keepaspectratio=true]{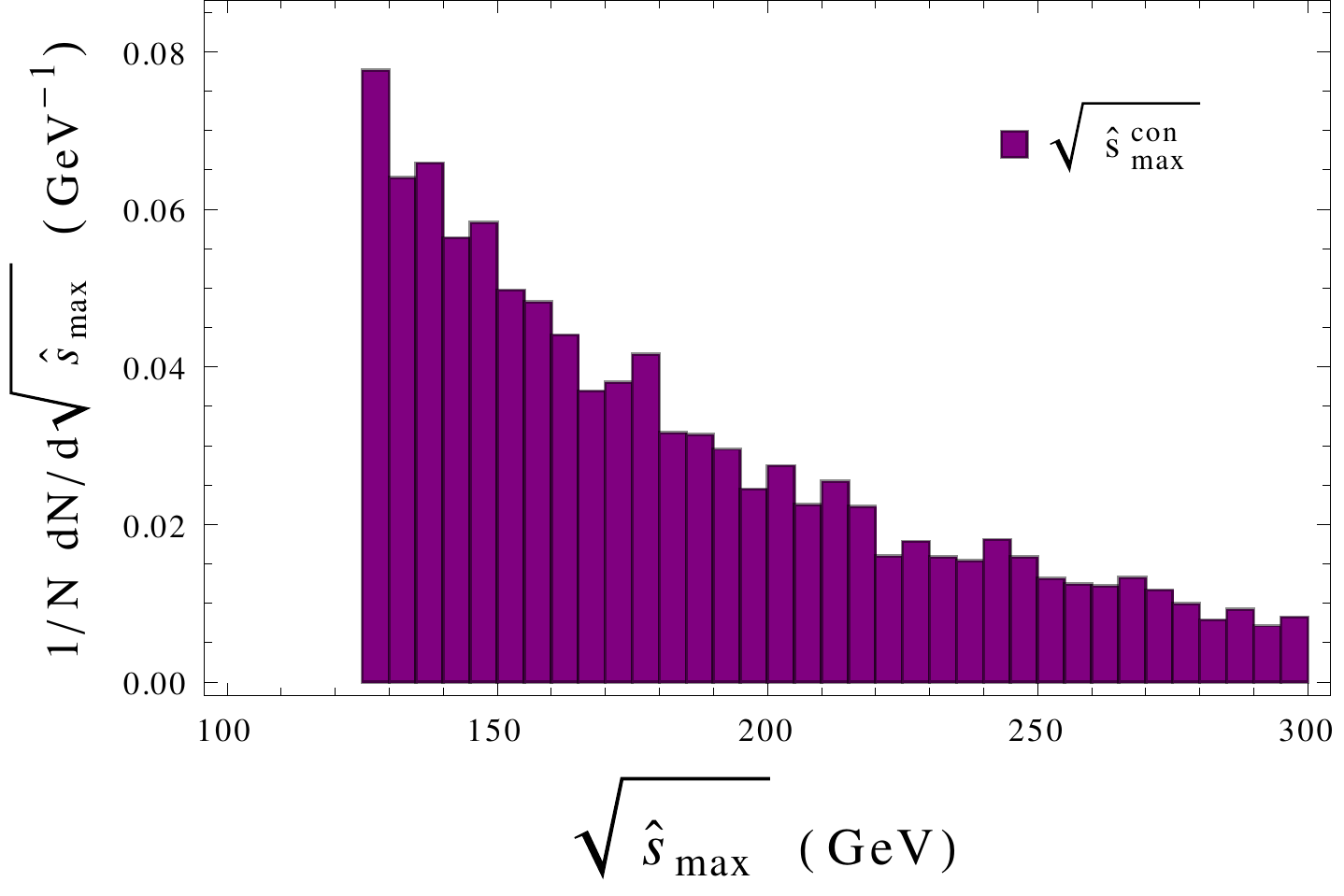}
  \caption{
 (Left) Figure shows the distribution of $\hat{s}_{min}$ and $\hat{s}_{min}^{cons}$ with $\hat{s}^{True} = M_h = 125.0$ GeV. The red colored histogram is for analytical formula of $\hat{s}_{min}$ which also can be verified using numerical  minimisation, blue colored histogram is $\hat{s}_{min}^{cons}$ calculated using numerical minimisation. The variables $\hat{s}_{min}$  and $\hat{s}_{min}^{cons}$ have endpoint at the heavy resonance mass $M_h$ but $\hat{s}_{min}^{cons}$ have larger number of events because of extra constraints it uses in its minimisation.
 (Right) Figure shows the distribution of the $\hat{s}_{max}^{cons}$, as one can see it has threshold at the true mass of the heavy  resonance $M_h = 125.0$ GeV. As one can see the $\hat{s}_{max}^{cons}$ always greater than or equal to $\hat{s}^{True}$. Similar unconstrained variable {\it i.e.} $\hat{s}_{max}$ is not present.
 }
 \label{fig:shatconsminmax}
\end{figure}

 To illustrate the properties of these constraint variables, first we consider a simple example of SM 
 Higgs production through gluon fusion 
 at the hadron collider. Higgs boson decays further semi-invisibly through tau pair production, 
 $h \rightarrow \tau \tau \rightarrow w \nu_{\tau}  w \nu_{\tau}$. 
 To compare with the representative diagram for the antler production in Fig.~\ref{fig:AntlerTopology}, 
 $\tau$ being the intermediate particle $B_i$, for which additional mass-shell condition used in the minimisation(maximisation) 
 of constrained $\hat{s}$. Neutrino $\nu_{\tau}$ is the invisible particle $X_i$. We considered hadronic (leptonic)
 decay mode for the $W$ boson thereafter to consider two invisibles (four invisibles tested in next example) in the final state.
 The  distribution of $\hat{s}_{min}$ and $\hat{s}_{min}^{cons}$ are shown in the Fig.~\ref{fig:shatconsminmax}(left).
 Red binned histogram shows the distribution for $\hat{s}_{min}$, which can be calculated numerically or using analytical expression. 
 Blue histogram  shows the distribution of constrained $\hat{s}_{min}^{cons}$.  As expected the 
 the endpoint of both the $\hat{s}_{min}$  and $\hat{s}_{min}^{cons}$ distributions are at the $\sqrt{\hat{s}^{True}}=M_h=125\;GeV$
 for a choice of invisible mass as zero. 
 Evidently larger number of events at the endpoint for the $\hat{s}_{min}^{cons}$ distribution with a sharper drop can be
 measured more precisely. This is even more important once corresponding background also considered together. 
 The Fig.~\ref{fig:shatconsminmax}(right) demonstrates the distribution of other constraint variable $\hat{s}_{max}^{cons}$ 
 which has a threshold at $\hat{s}^{True}$ with  considerable number of events at the threshold.

 \begin{figure}[t]
 \centering
 \includegraphics[bb=300 260 566 353,scale=0.52,keepaspectratio=true]{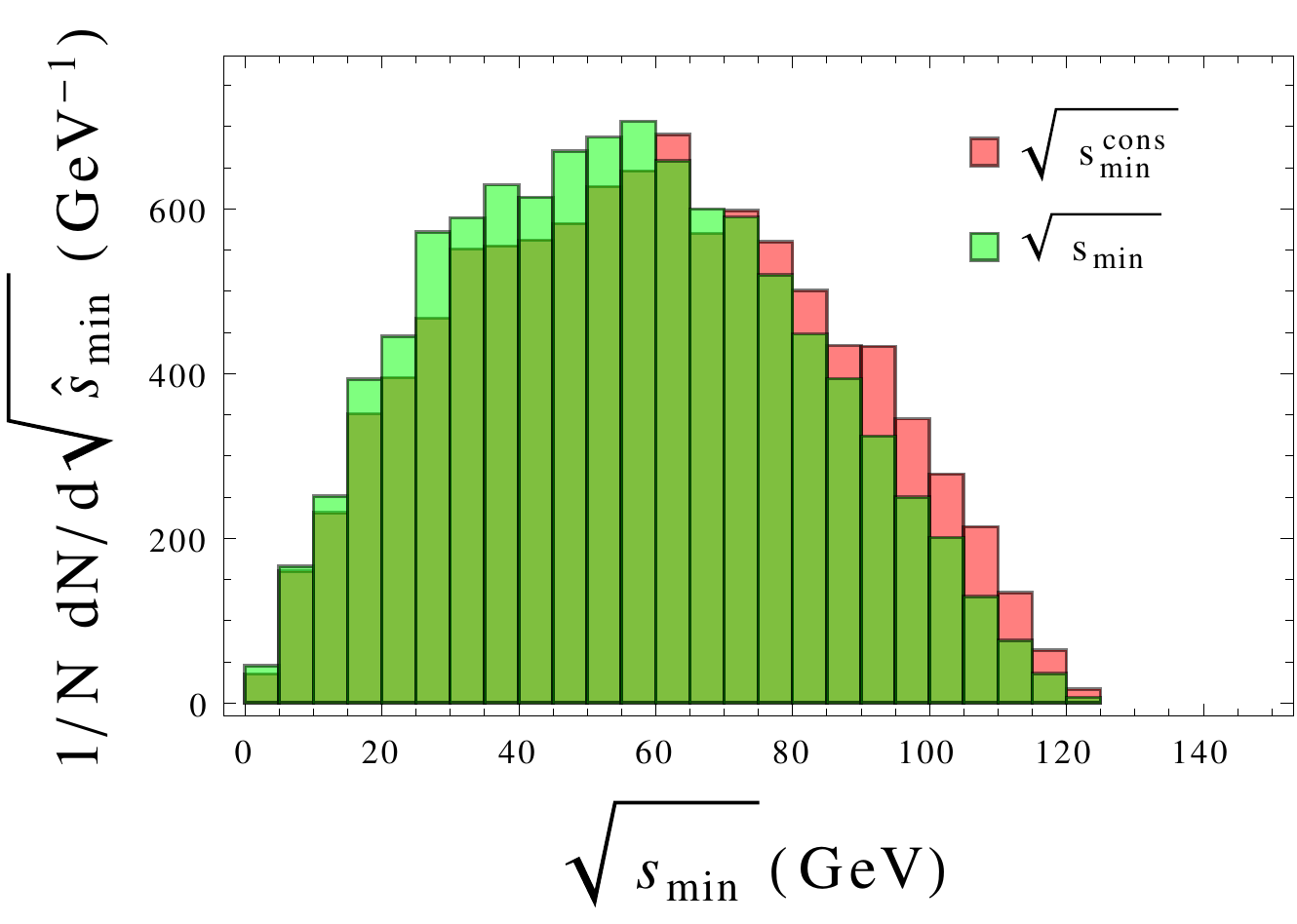}
 \hskip 0.2cm
 \includegraphics[bb=-400 0 507 282,scale=0.52,keepaspectratio=true]{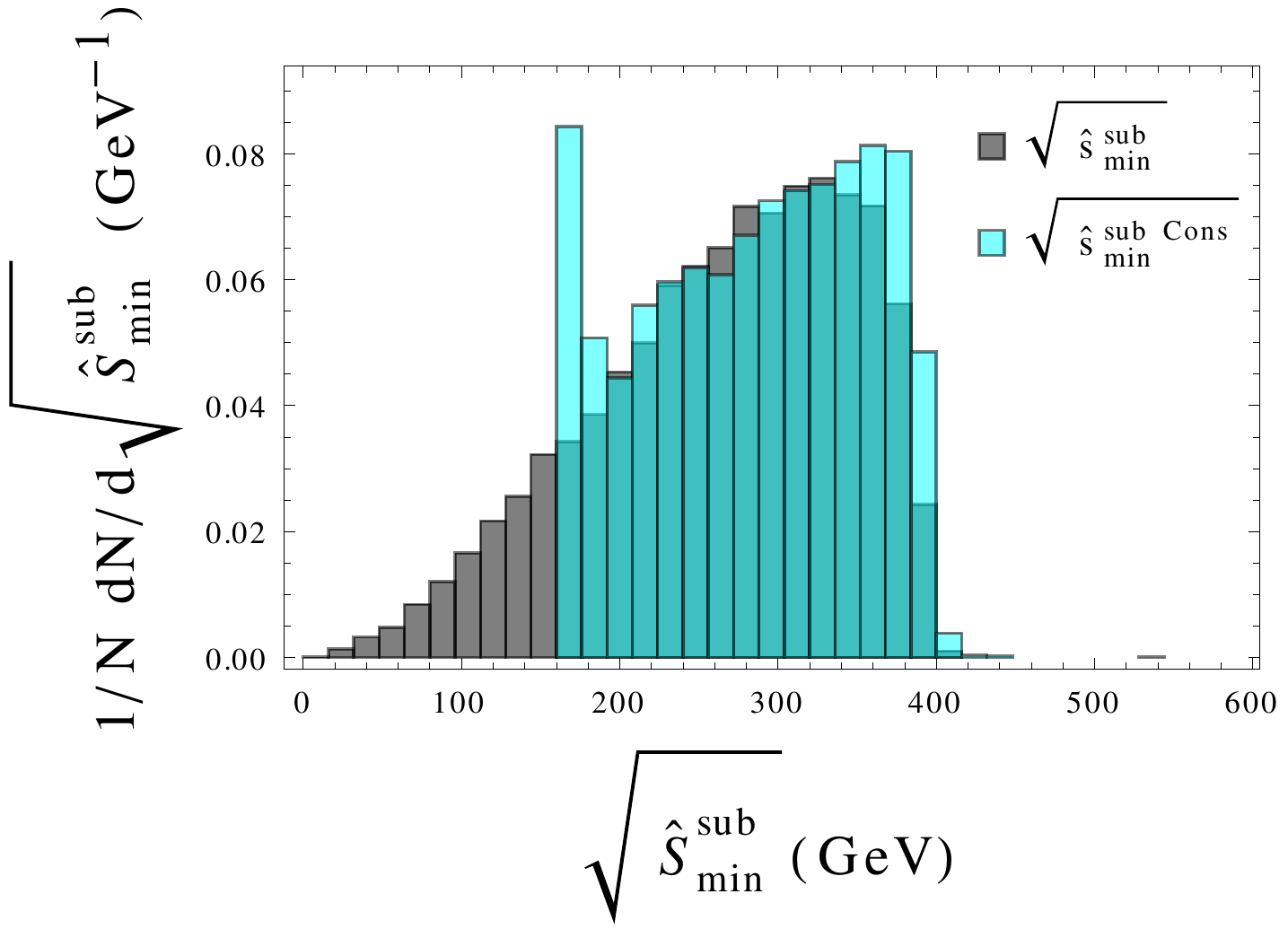}
  \caption{
  (Left) In this figure we have shown the distribution of the $\hat{s}_{min}$ and $\hat{s}_{min}^{cons}$ for for four invisible particle in the  final state. The red colored histogram shows the distribution of $\hat{s}_{min}^{cons}$ and the green colored histogram shows the distribution $\hat{s}_{min}$. As one can see though there are endpoint feature for both the variables but the number events is very less and the improvement for $\hat{s}_{min}^{cons}$ over $\hat{s}_{min}$ is also very less.
  (Right) Here the red colored histogram is $\hat{s}_{min}^{sub}$ calculated using analytical formula, the blue colored histogram is
 $\hat{s}_{min}^{sub}$ calculated numerically and the green colored histogram is $\hat{s}_{min}^{sub, cons}$ calculated numerically.
 The variable $\hat{s}_{min}^{sub}$ and $\hat{s}_{min}^{sub, cons}$ has endpoint at  $\hat{s}^{sub, True} = M_{\phi^{++}}$.
  }
\label{fig:multiInvisibleFinalState_phidoublePlus}
\end{figure}

  It is expected that the endpoint of kinematic distribution would be less populated if one has more number of 
  invisibles in the event. This is because increasing number of invisibles would increase the number of unknown 
  momentums restricted with the same constraints.
  Following our previous example, we now consider the four invisible particle by decaying both $w$ leptonically 
  and demonstrated corresponding $\hat{s}_{min}$ and $\hat{s}_{min}^{cons}$ distributions in  Fig.~\ref{fig:multiInvisibleFinalState_phidoublePlus}(left). 
  The red histogram  shows the distribution for $\hat{s}_{min}^{cons}$, whereas the green binned histogram shows $\hat{s}_{min}$ 
  to compare the effect due to the extra constraints. These distribution confirms that  the number of events at the endpoint 
  are considerably low as one increases the number of invisible particle in the final state. Although
  constrained variable can improve the situation only slightly, overall both these distributions forms a narrow tail 
  rather than the sharp endpoint.
  
  We further study one more interesting example from the resonant production of exotic doubly charged 
  scalar~\cite{Bambhaniya:2013yca} production at the hadron collider following its decay into dominant decay channel
  producing pair of $w$, which in turn decays leptonically. Hence, the resonant sub-system under consideration 
  is $\phi^{++} \rightarrow w^{+} w^{+} \rightarrow \ell^{+} \nu_{\ell} \ell^{+} \nu_{\ell}$. 
  In hadron collider this exotic state $\phi^{++}$ can 
  be produced associated with charged $w^{-}$ which mainly decays hadronically and it is possible to disentangle 
  from the antler sub-system producing lepton pair from the exotic decay. 
  We choose to use corresponding subsystem variable $\hat{s}_{min}^{sub}$ for our analysis. Here analytical expressions 
  for the invisible particle momenta remain same except the modified form for $\mptvec$ which includes the visible contribution
  from non-sub-system~\cite{Konar:2010ma}. 
  The distributions for the $\hat{s}_{min}^{sub}$ and the constrained variable $\hat{s}_{min}^{sub, cons}$ are demonstrated in 
  the Fig.~\ref{fig:multiInvisibleFinalState_phidoublePlus}(right). 
  Dark binned histogram represents the distribution for $\hat{s}_{min}^{sub}$ which can be calculated both analytically or 
  using numerical minimisation. The cyan colored histogram is the distribution for $\hat{s}_{min}^{sub, cons}$
  utilising extra $w$ mass-shell constrains, and minimised numerically. 
  One can note that the $\hat{s}_{min}^{sub, cons}$ is performing better in getting  the endpoint at $\phi^{++}$ mass.
  Observed small tail is because of finite width from $\phi^{++}$ and these extra constraints ensures that the 
  $\hat{s}_{min}^{sub, cons}$ distribution starts from a threshold at the scale of $2m_w$.


 \section{Non-antler topology and constrained variables}\label{sec:4}

   Non-antler topology is extremely common in most of the BSM theories and also abundant in SM. This topology is already described
   in the Fig.~\ref{fig:NonAntlerTopology}, where $B_i$ are the parent particles produced in pair. After a cascade of decay each side of 
   the decay chain produces number of visibles along with a massive invisible particle $X_i$.  Detailed discussion on the 
   behaviour of $\hat{s}_{min}$ as a mass bound variable is done extensively for this kind of topology. Here we would illustrate
   the constrained variables in the light of additional on-shell constraints. 
   Unlike using these exact on-shell constraints for the parents mass, which is primarily one would like to know through these 
   mass bound variables, ref.~\cite{Cho:2014naa} uses constraints from the equality of two parents mass. Following our analysis 
   in previous section, we continue using these mass-shell constraints with an expectation of improved momentum reconstruction. 

 \begin{figure}[t]
 \centering
 \includegraphics[bb=0 0 601 371,scale=0.5,keepaspectratio=true]{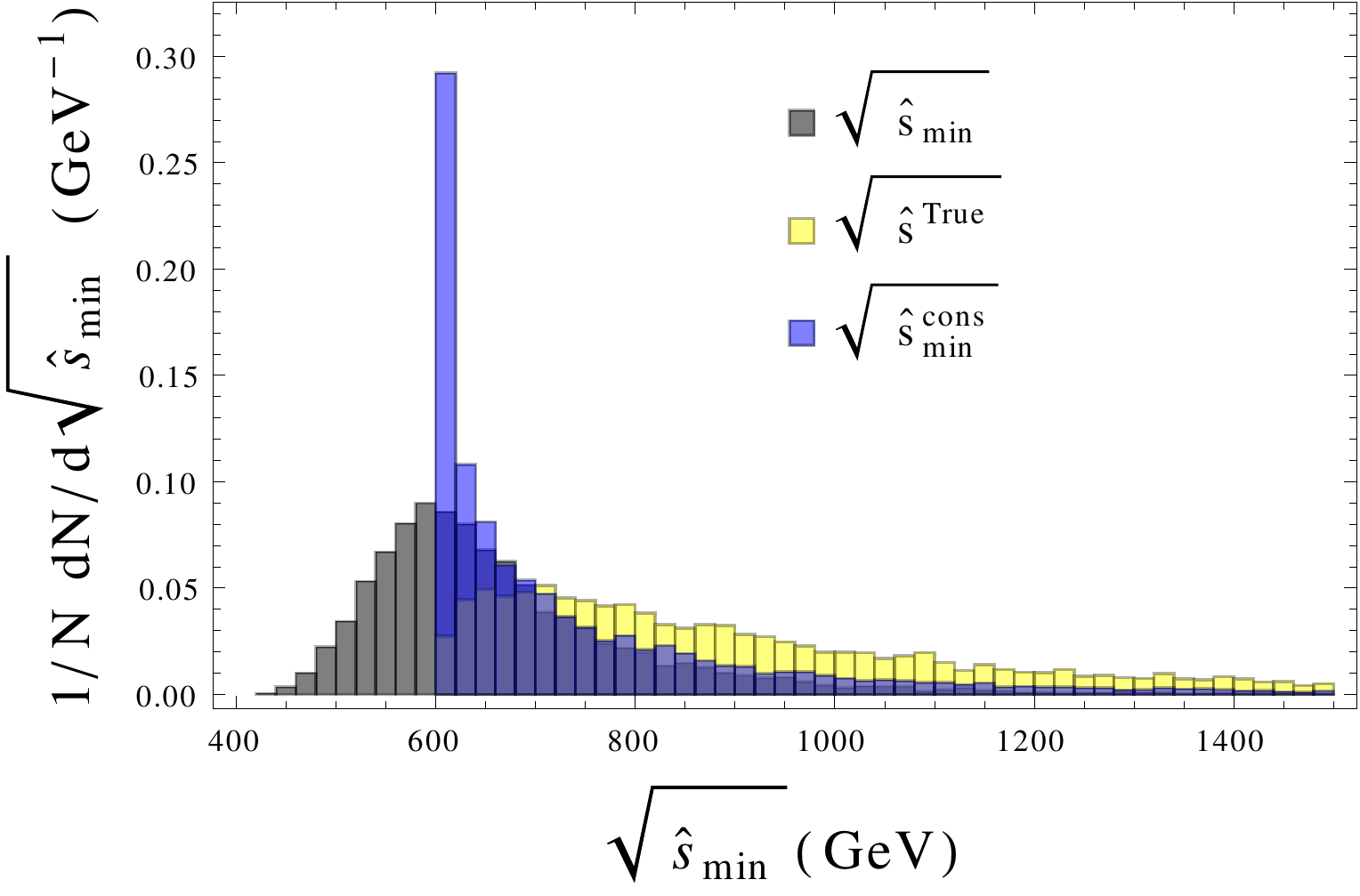}
\caption{Figure shows the distribution of $\hat{s}_{min}$ (black) and $\hat{s}_{min}^{cons}$ (blue) considering a
toy model of non-antler pair production at the the hadron collider, with parent and invisible mass as 
300 GeV and 200 GeV respectively.
Non-antler heavy parent particle pair production must have a true parton level CM energy distribution starting 
from a threshold value of total parents mass as shown by yellow colored histogram.
As a consequence of additional constraints $\hat{s}_{min}^{cons}$ distribution also poses this same threshold,
however with a considerable number of events at the threshold.
}
 \label{fig:constrainedShatForNonAntlerTopology}
\end{figure} 
   
   As a consequence of on-shell constraints, one can expect $\hat{s}_{min}^{cons}$ distribution would start from
   a threshold at the sum of parents mass. This is contrary to the unconstrained $\hat{s}_{min}$ distribution which 
   exhibits peak at that position giving  an excellent correlation for the new physics mass scale.
   This is demonstrated in Fig.~\ref{fig:constrainedShatForNonAntlerTopology} where distributions for $\hat{s}_{min}$ (black) 
   and $\hat{s}_{min}^{cons}$ (blue) are plotted using a toy model of non-antler pair production at the the hadron collider, 
   with parent and invisible mass as 300 GeV and 200 GeV respectively.
   Unlike antler decay topology where heavy particle resonant production form a near delta function at the parton level
   center of mass energy, here heavy parent particle pair production has a distribution starting from a threshold value of total 
   parents mass as shown by yellow colored histogram.
   As we note that the $\hat{s}_{min}^{cons}$ distribution also poses this same threshold,
   however with a considerable number of events at the threshold.
   We will follow further in the next section to show the improvements in invisible momentum construction as a presence 
   of these constraints and the events contributing at the threshold.
   Analogous to the variables constructed for antler topology, one can follow the similar hierarchy among all the 
   constrained $\hat{s}$ mass variables after imposing different constrains:
 \begin{equation}
  \hat{s}_{min} \le \hat{s}_{min}^{cons} \le \hat{s}^{True} \le \hat{s}_{max}^{cons}.
 \end{equation}

 \begin{figure}[t]
 \centering
 \includegraphics[bb=0 0 360 349,scale=0.5,keepaspectratio=true]{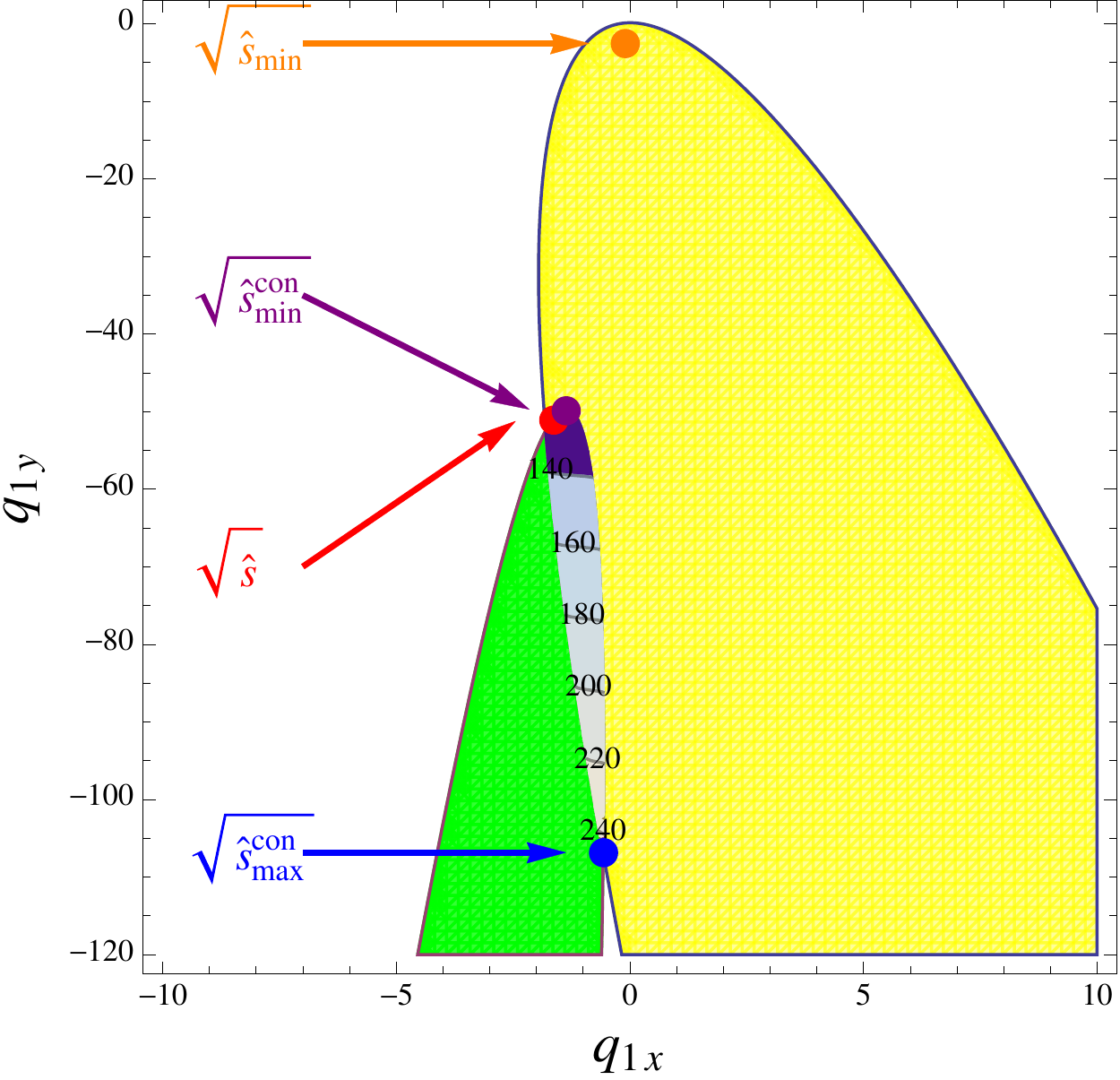}
 \hskip 1cm
 \includegraphics[bb=0 0 360 346,scale=0.5,keepaspectratio=true]{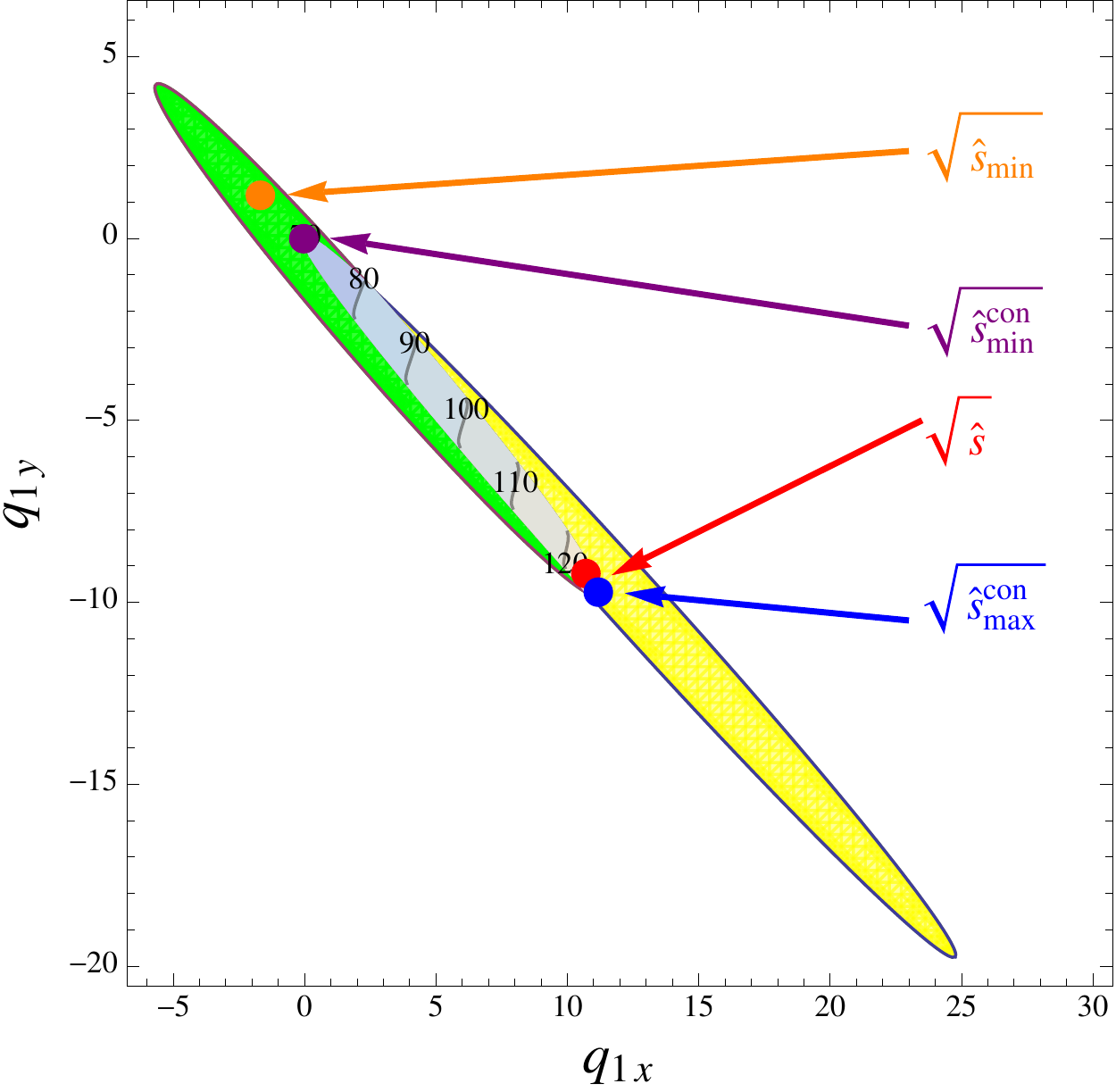}
 \caption
 {
 Some example events demonstrating the invisible momentum reconstruction in case of antler topology 
 through minimisation during construction of 
 different $\hat{s}$ variables. Each color shaded region is representing the allowed phase space by additional constrains
 in the unknown invisible momentum parameter space.
 In both of the plot yellow elliptical region is constrained area describing $q_{1z}(\vec{q}_{1T})$ and 
 green elliptical region is constrained area for $q_{2z}(\vec{q}_{1T})$. Intersection region between these 
 two constrained ellipse, shaded in white, is eligible for containing all the constraint $\hat{s}(\vec{q}_{1T})$ 
 parameters as well as the true $\hat{s}$. 
 Two other ends in this overlapping region would typically represent the $\hat{s}_{min}^{cons}$ and $\hat{s}_{max}^{cons}$,
 with true $\hat{s}$ in between them.
 Since $\hat{s}_{min}$ does not have this additional constrain, it would be outside overlapping region and far from true $\hat{s}$.
 Inside the overlapping region $\hat{s}$ contours are also presented where true CM energy is matches with the value of Higgs mass.
 The left figure shows one example event where $\hat{s}^{True}$ is closer to $\hat{s}_{min}^{cons}$. This contributes 
 at the endpoint of the $\hat{s}_{min}^{cons}$ distribution and also giving better momentum reconstruction.
 The right figure shows another event where $\hat{s}^{True}$ is close to the $\hat{s}_{max}^{cons}$ contributing at the 
 threshold of this distribution with better momentum reconstruction.
 }
 \label{fig:shatTrueclosetoshatconsmin}
\end{figure}

 \section{Event reconstruction capability}\label{sec:5}

 In this section we describe the invisible momentum reconstruction capability using mass variables 
 $\hat{s}_{min}$ and improvement in it accounting for additional constraints in the context of antler and non-antler
 decay topology. Analytic expressions for invisible momenta components from the $\hat{s}_{min}$ was already discussed 
 in Sec.~\ref{sec:2}. It was also argued that these reconstructed invisible momenta using $\hat{s}_{min}$ are unique 
 irrespective of any topology considered.  
 Note that these reconstructed momenta from the minimisation of $\hat{s}_{min}$ are not the true momenta, 
 but approximated momenta consistent with the observables in such event. These calculated momenta can be 
 correlated with the true values of them to find the reconstruction efficiency similar to the other reconstruction methods 
 like MAOS~\cite{MAOS1,MAOSclassification}.

 To describe the consequence of the constraints given in Eqs.~\ref{mother1cons}-\ref{daughter2cons} in constructing the 
 new variables $\hat{s}_{min}^{cons}$  and $\hat{s}_{max}^{cons}$, we reorient them to write unknown 
 longitudinal momenta in terms of their transverse components $\vec{q}_{iT}$. We get,
  \begin{equation}\label{longitudinal}
  q_{iz} = \frac{\Sigma_i P_{iz}^V \pm E_i^V \sqrt{\Sigma_i^2 - (E_{iT}^V E_{iT}^q)^2}}{(E_{iT}^V)^2} \, ,
 \end{equation}
 with
 \begin{eqnarray}
  \Sigma_i &=& \frac{M_{Bi}^2 - M_{Xi}^2 - M_{vi}^2}{2} + \vec{P}_{iT}^V.\vec{q}_{iT} \, ,\\
  E_{iT}^V &=& \sqrt{M_{vi}^2 + (p_{iT}^V)^2} \, , \\
  E_{iT}^q &=& \sqrt{M_{Xi}^2 + q_{iT}^2}
 \end{eqnarray}
 where $M_{vi}$ is the invariant mass of visibles in the $i$-th decay chain, $i = 1, 2$. 
 Missing transverse  momentum constraints further permit us to rewrite them in terms of a single 
 invisible particle transverse momentum components, which we choose as $\vec{q}_{1T}$ for our examples.
 By simplifying the right hand side of the Eq.~\ref{longitudinal}, one gets the equation of 
 ellipse in terms of the transverse momenta and the parameters outside the ellipse are not physical
 with the given event. Two elliptical allowed regions for each event correspond to two side of decay
 chain and these two regions can not be completely disjoint from each other. 
 All the available constraints in an event are satisfied only at the intersection region between them. 
 Different situations can emerge for this overlapping region. Two ellipses may intersect each other over a
 finite region or a point (touching each other). In other case  one ellipse may contain the other ellipse.
 
\begin{figure}[t]
 \centering
 \includegraphics[bb=0 0 360 341,scale=0.5,keepaspectratio=true]{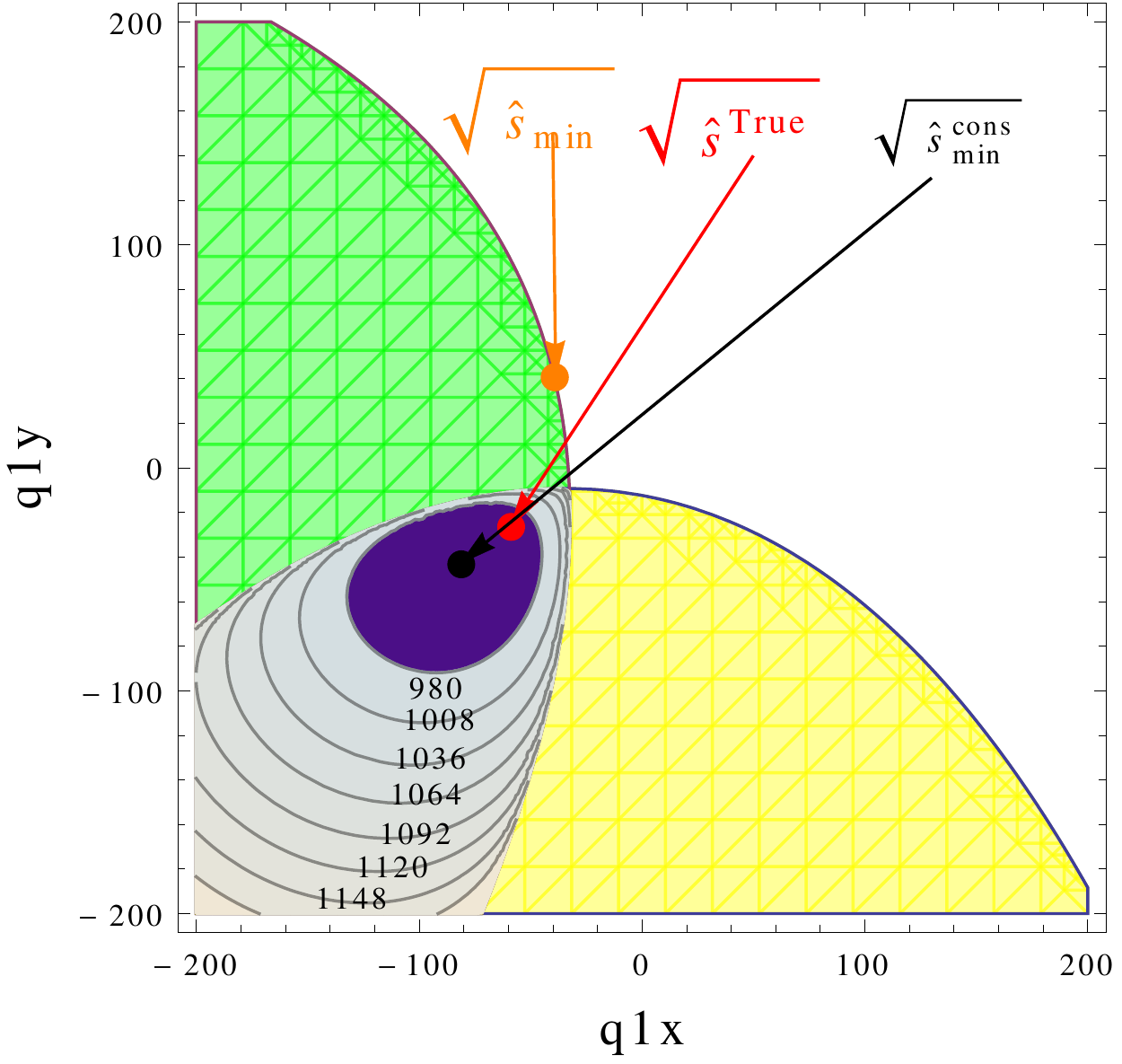}
 \caption{
 One example event demonstrating the invisible momentum reconstruction in case of non-antler topology 
 through minimisation during construction of different $\hat{s}$ variables. Description of  shaded regions and mass variables
 are similar to previous figure.
 }
 \label{fig:MomentumReconsNonAntler}
\end{figure}

 In Fig.~\ref{fig:shatTrueclosetoshatconsmin} we consider such constrained regions demonstrated for two different 
 events in antler topology. Each color shaded region is representing the allowed phase space by additional constrains in the unknown 
 invisible momentum parameter space. Overlapping region between these two constrained ellipse is shaded in white where
 $\hat{s}$ contours are also presented. One can identify the minimum value from this intersection region as
 $\hat{s}_{min}^{cons}$ and the maximum value as $\hat{s}_{max}^{cons}$ which reside at opposite ends within this region.
 Since $\hat{s}^{True}$ also satisfies all the constraints in the event it must also remain in the intersection region and
 in between these those two constrained points\footnote{Eq.~\ref{longitudinal} reflects a four fold ambiguity from the 
 longitudinal component in each event. However, the extremisation of constrained $\hat{s}$ would qualify for a choice of unique 
 momentum reconstruction.}.
 Since the $\hat{s}_{min}$ variable does not satisfy all additional constraints in the event, it would lie outside the 
 intersection region and relatively far from true value.
 The left figure display one typical example event where $\hat{s}^{True}$ is closer to $\hat{s}_{min}^{cons}$. This contributes 
 at the endpoint of the $\hat{s}_{min}^{cons}$ distribution and also giving better momentum reconstruction. 
 The right figure shows another event where $\hat{s}^{True}$ is close to the $\hat{s}_{max}^{cons}$ contributing at the 
 threshold of this distribution with better momentum reconstruction. In both figure, we depicted different colored dots for
 the the position (invisible momentums during minimusation or maximisation) of all $\hat{s}$ variables together with actual 
 $\hat{s}$ correspond to that particular event. One can even read the corresponding values of these mass variables from their 
 contours plotted within intersecting region.
 Similarly in Fig.~\ref{fig:MomentumReconsNonAntler} we have shown the momentum reconstruction capability of 
 $\hat{s}_{min}^{cons}$ and  $\hat{s}_{min}$ in an example of Non-Antler topology. The yellow and green shaded regions 
 represents constrained $q_{1z}(\vec{q}_{1T})$ and $q_{2z}(\vec{q}_{1T})$ respectively and their intersection 
 region is suitable for constrained $\hat{s}$. 
 The red, orange and black point shows the true momenta and reconstructed momenta given by $\hat{s}_{min}$,
 $\hat{s}_{min}^{cons}$ respectively.

 \begin{figure}[t]
 \centering
 \includegraphics[bb=0 0 360 349,scale=0.55,keepaspectratio=true]{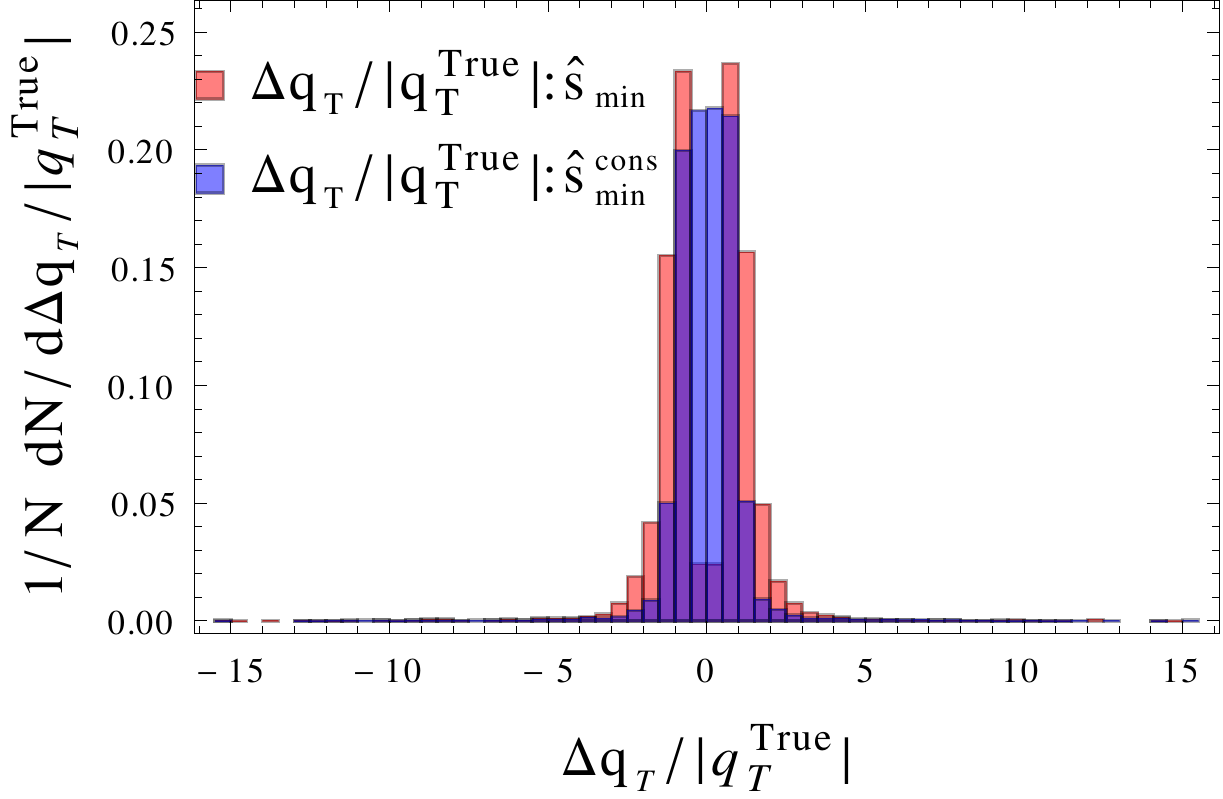}
 \hskip 0.9cm
 \includegraphics[bb=0 0 360 346,scale=0.55,keepaspectratio=true]{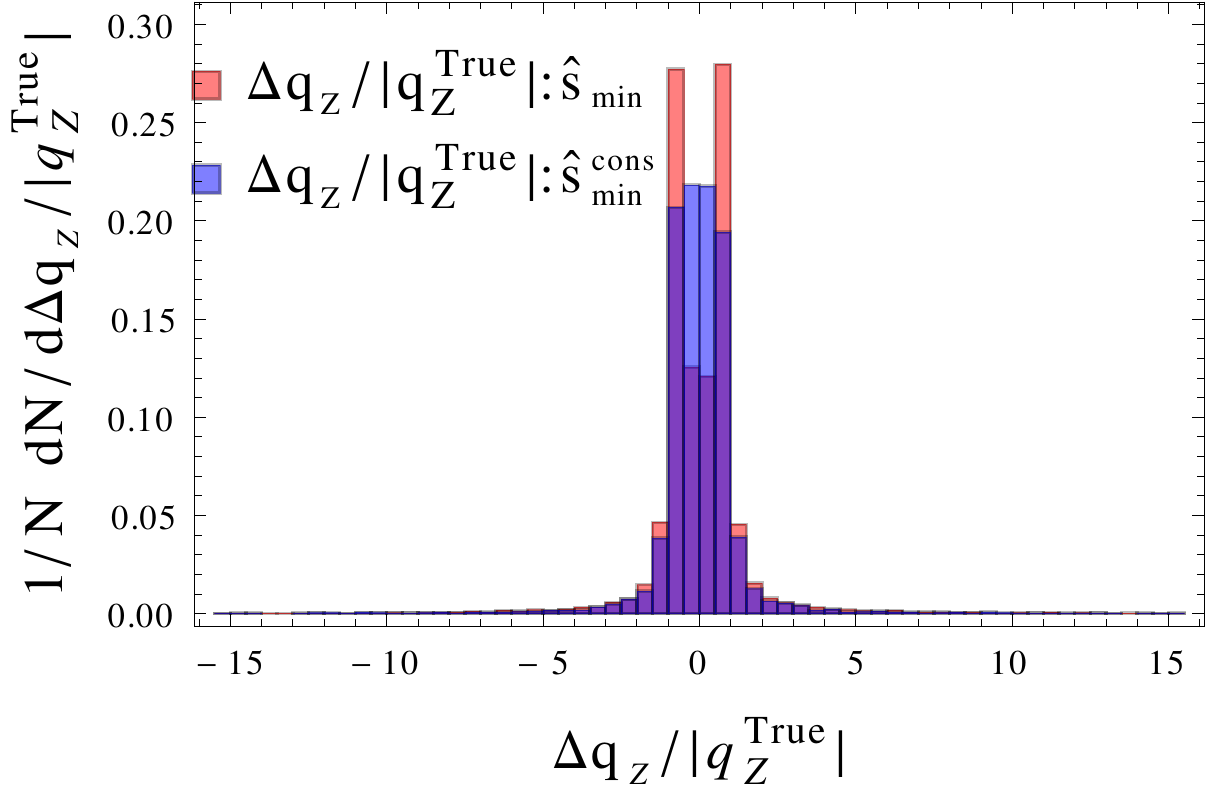}
 \caption{
 Histogram showing the distributions for deviation of the reconstructed momentum from the corresponding true 
 momentum as a fraction of true momentum $(q_i^{reconstructed}-q_i^{true})/|q_i^{true}|$ using both unconstrained (red)
 and constrained (blue) $\hat{s}_{min}$ methods.
 Left (right) plot displays the momentum reconstruction capability in antler topology for transverse (longitudinal) 
 components of momentum. In each figure shown histograms one directly from $\hat{s}_{min}$ for which agrees with 
 corresponding analytical form, also shown histograms using constrained minimisation $\hat{s}_{min}^{con}$ to compare 
 the improvements over unconstrained ones.}
 \label{fig:reco_mom_dist}
\end{figure}

 We are now in a position to quantify the capability of momentum reconstruction. Fig.~\ref{fig:reco_mom_dist} exhibits
 the histogram showing the distributions for deviation of the reconstructed momentum from the corresponding true 
 momentum as a fraction of true momentum $(q_i^{reconstructed}-q_i^{true})/|q_i^{true}|$ using both unconstrained 
 and constrained  $\hat{s}_{min}$ methods.
 Left plot displays the momentum reconstruction capability in antler topology for transverse 
 components of momentum. Similarly, right plot is for corresponding longitudinal component of the momentum. 
 In each figure one histogram (in red bins) is shown for $\hat{s}_{min}$ which agrees with 
 corresponding analytical form. Also histograms with blue bins plotted in the same figure to display the
 momentum reconstruction capability using constrained minimisation $\hat{s}_{min}^{cons}$ pointing out
 improvements over unconstrained ones.

\begin{figure}[t]
 \centering
 \includegraphics[bb=0 0 378 243,scale=0.55,keepaspectratio=true]{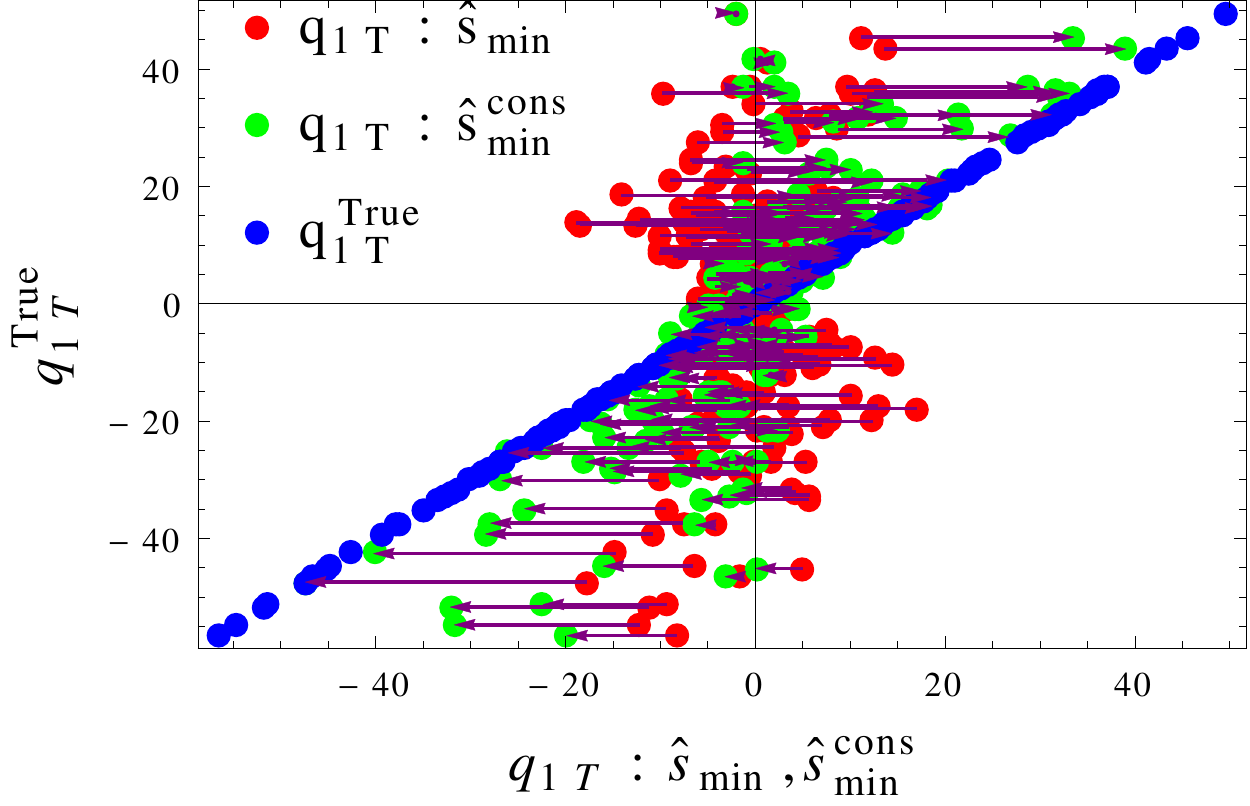}
 \hskip 0.7 cm
 \includegraphics[bb=0 0 360 228,scale=0.55,keepaspectratio=true]{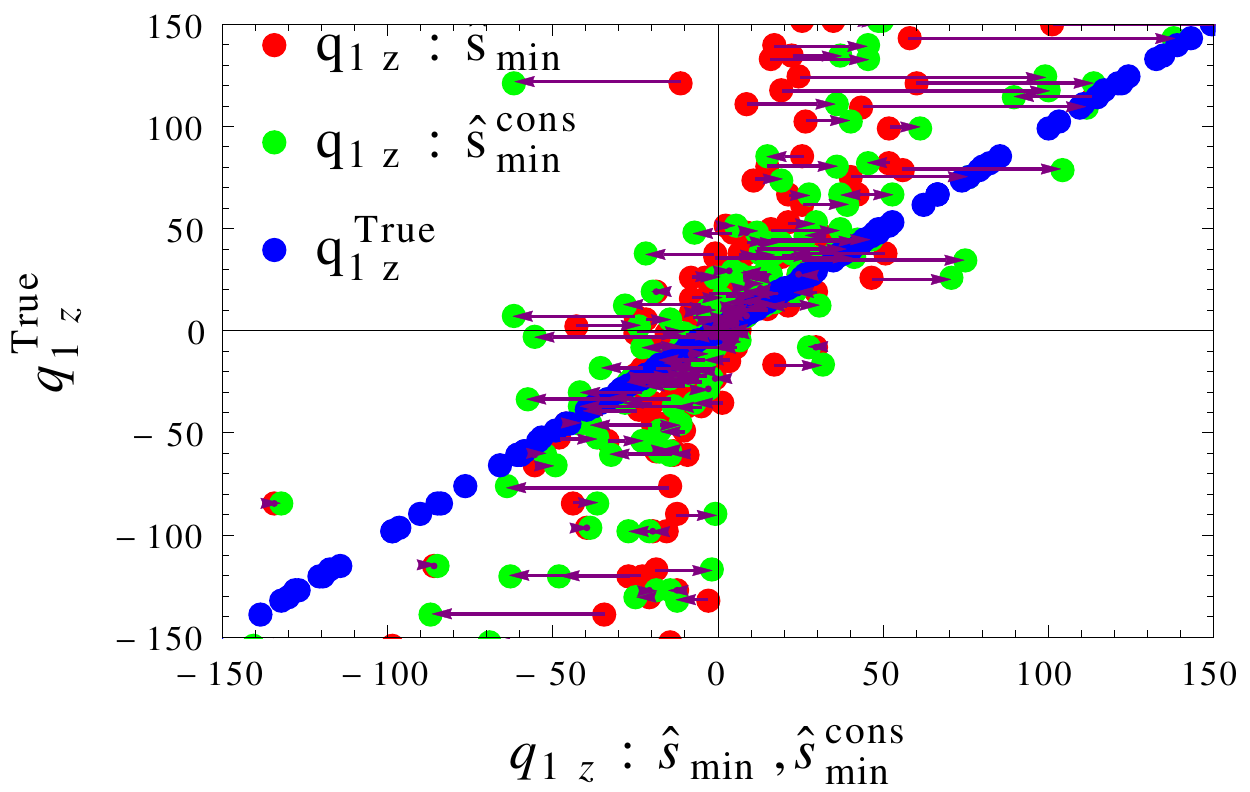}
 
 \caption{Correlation plot taking few random representative event points showing the shift of reconstructed transverse momenta (in left plot) and longitudinal momenta (in right plot) derived from $\hat{s}_{min}$ and $\hat{s}_{min}^{cons}$. In both plots, each red dot point represents the calculated momentum derived from the $\hat{s}_{min}$ against the corresponding true momentum for each event. Similarly, green dots are for corresponding momentum derived from the $\hat{s}_{min}^{cons}$.  The purple arrows connecting from one red dot to other green dot represent the shift in the derived momentum once extra constraints are imposed. Diagonal blue points the simply correlate true momenta with true momenta in each event to give the perspective how derived momenta composed against the true values. 
}
 \label{fig:shiftshatminVsConshatminq1T}
\end{figure}

 We discussed the additional constraints in $\hat{s}_{min}$ to choose the minimisation 
 that gives reconstructed invisible momenta closer to their true values.
 To understand this consequence better, we look into the movements of these calculated momenta
 once we impose the constraints.  In Fig.~\ref{fig:shiftshatminVsConshatminq1T}
 we demonstrate through a correlation plot of constructed invisible momentum versus the corresponding true
 momentum taking few random representative event points.
 In both plots, each red dot point represents the calculated momentum derived from the $\hat{s}_{min}$ against 
 the corresponding true momentum for each event. Similarly, green dots are for corresponding momentum derived 
 from the $\hat{s}_{min}^{cons}$.  The purple arrows connecting from one red dot to other green dot represent 
 the shift in the derived momentum once extra constraints are imposed. 
 Since the true momentum is always same for a particular event, shifts due to minimisation in different mass 
 variables are only horizontal. These arrows represent the degree of change due to constraints, shifting calculated 
 momentums towards the diagonal true momentum points.
 Diagonal blue points the simply correlate true momenta with true momenta in each event to give the perspective 
 how derived momenta composed against the true values.
 Left (right) plot corresponds to the transverse (longitudinal) momenta derived from $\hat{s}_{min}$ and $\hat{s}_{min}^{cons}$.
 
%

 \section{Summary and conclusions} \label{sec:summary}

Large Hadron Collider started its extremely successful journey finding the long sought scalar. 
With no substantial evidence for BSM, expectation is high for next run of LHC. In the light of 
dark matter models, missing energy signals would be looked very carefully. The $\hat{s}_{min}$ 
variants of mass variables were designed for prompt finding of mass scale in a model independent 
way for any complex topology of BSM events associated with semi-invisible final production. 
In the present analysis, we proposed to exploit additional partial informations available in the event
as constraints to improve the finding. 
We classified our discussions based on two different class of simple production topology widely available
both in SM and BSM production, which are, antler and non-antler. 

Different SM as well as new physics predicts antler production processes, including important Higgs 
production in the hadron collider. These topology can be constrained significantly using additional 
intermediate mass-shell conditions. We have demonstrated with different examples to show 
that the constrained variable $\hat{s}^{cons}_{min}$ can significantly improve the distribution 
and the measurements. More interestingly, these additional constraints ensures a finite upper value 
of the $\hat{s}$ variable, defined as, $\hat{s}^{cons}_{max}$ which is not well-defined and finite 
in the unconstrained picture. Hence, this new variable also can be exploited up to some extend. 
Apart from considering different BSM example to demonstrate these variable in the context of sub-system
topology and in the difficult signatures with more invisible final states in antler topology,
we also demonstrated effect of these additional constraints in a simple non-antler topology. 

To clarify the effects of these constraints in the invisible momenta parameter space, we choose
phenomenological examples explicitly demonstrating how these mass variables are restricted and 
pushed towards the true values of $\hat{s}$, together with their choice the 
invisible momentums closer to that of true one. Hence, one can consider to quantify the capability of 
reconstructing the invisible momenta in present scenario. 
We constructed and shown the efficiency of momentum reconstruction using these constrained $\hat{s}$ variables which 
predicts a unique momenta associated with each of these mass-bound variables in each event. 
In conclusion, we explored the utility of additional informations we may already have, during exploration of the
$\hat{s}$ type mass variables which can be exploited with missing energy data at the LHC.

\bigskip
\acknowledgments
We would like to thank A. Datta for useful discussion during WHEPP XIII. P.K. also thanks RECAPP, 
HRI for hospitality where part of this work has been carried out.
\listoftables           
\listoffigures          

\bibliographystyle{unsrt}
\bibliography{bibliography}

\end{document}